# The Influence of Electric Field on the Anisotropic Dispersion of the Flexocoupling Induced Phonons and Ferrons in Van der Waals Ferrielectrics

Anna N. Morozovska[1,*], Eugene A. Eliseev[2], Yujie Zhu[3], Yulian M. Vysochanskii[4], Venkatraman Gopalan[5], Long-Qing Chen[5,†], and Jia-Mian Hu[3,‡]

[1] Institute of Physics, National Academy of Sciences of Ukraine, 46, pr. Nauky, 03028 Kyiv, Ukraine

[2] Frantsevich Institute for Problems in Materials Science, National Academy of Sciences of Ukraine, Omeliana Pritsaka str., 3, Kyiv, 03142, Ukraine

[3] Department of Materials Science and Engineering, University of Wisconsin-Madison, Madison, WI, 53706, USA

[4] Institute of Solid-State Physics and Chemistry, Uzhhorod University, 88000 Uzhhorod, Ukraine

[5] Department of Materials Science and Engineering, Pennsylvania State University, University Park, PA 16802, USA

**Abstract**

As has been shown recently, the influence of the flexoelectric coupling (shortly "*flexocoupling*") on the fluctuations of electric polarization and elastic strains can lead to the principal changes of the dispersion law of soft optical and acoustic phonons (shortly "*flexophonons*") and ferrons (shortly "*flexoferrons*") in van der Waals ferrielectrics. Analytical results, derived in the framework of Landau-Ginzburg-Devonshire approach, revealed that the dispersion of flexophonons and flexoferrons is strongly anisotropic and should depend on the magnitude and direction of applied electric field. In this work we study the influence of applied electric field on the anisotropic dispersion of the flexophonons and flexophonons in a uniaxial van der Waals ferrielectric $CuInP_2S_6$. We reveal that the frequency of acoustic flexophonons and flexoferrons tends to zero at nonzero wavevectors under increase of applied electric field. We relate the changes with a possible appearance of a spatially modulated incommensurate polar phase induced by the flexocoupling in external field. The critical strength of flexocoupling is determined by the magnitude of electric field and direction of the wavevector. This allows us to propose a method for estimating the strength of flexocoupling in van der Waals ferrielectrics, providing that the frequency of the acoustic flexophonon is zeroing at the threshold value of the electric field. Since the flexoelectric coefficients are poorly known in van der Waals ferrielectrics, obtained analytical results can be useful for their flexo-engineering.

---

[*] corresponding author, e-mail: anna.n.morozovska@gmail.com

[†] corresponding author, e-mail: lqc3@psu.edu

[‡] corresponding author, e-mail: jhu238@wisc.edu

## 1. Introduction

The flexoelectric effect [1, 2] can influence significantly the electric polarization and mechanical state of mesoscale and nanoscale ferroics with long-range ordered phases [3, 4], for which the strain and/or stress gradients are very strong at the surfaces, interfaces, around point and topological defects [5]. The flexoelectric coupling (shortly "*flexocoupling*") between the polarization and strain gradients [6] can be the reason of domain walls conductivity [7] and polarization of twin walls [8] in antiferrodistortive ferroelastics. Moreover, the flexo-antiferrodistortive coupling can induce the spatially modulated phases in a wide class of ferroics [9]. When the strength of the static flexocoupling overcomes the upper limit [10], a spatially modulated state of the polarization fluctuations should emerge [11].

Remarkedly, experimental measurements and theoretical calculations of soft phonon dispersion can be very informative to study the influence of static and dynamic flexocoupling on the spatially modulated long-range ordered phases in ferroelectrics. However, most neutron inelastic scattering [12, 13, 14] and Raman scattering [15, 16, 17] experimental results focus on material-specific features of the scattering spectra. The influence of static and dynamic flexocoupling on the soft optical and acoustic phonon dispersions in the long-range ordered phases of ferroelectrics were studied using Landau-Ginzburg-Devonshire (LGD) approach in Refs. [18, 19]. It was shown analytically that the dispersion of the acoustic phonons is much more sensitive to the magnitude of static flexocoupling than that of soft optical phonons. The frequency of acoustic phonons approaches zero and becomes purely imaginary when the constant of static flexocoupling exceeds a critical value, which depends on the electrostriction coupling and elastic compliances, temperature, second-order and higher-order polarization and strain gradient coefficients in the LGD free energy functional [11]. When the frequency of the acoustic phonons is zeroing, a spatially modulated incommensurate phase may appear in commensurate ferroelectrics [19]. The theoretical prediction [19] has found its experimental validation quite recently, when Orenstein et al. [20] observed polarization density waves in $SrTiO_3$ and attributed them with the flexocoupling, at that the flexocoupling strength was treated as the fitting parameter.

The concept of a "ferron" quasiparticle was introduced by Bauer et al. [21, 22] and Tang et al. [23]. They considered bosonic excitations in displacive ferroelectrics, which carry electric dipoles, from the Landau-type approach and introduced the ferrons emerging from the joint action of anharmonicity and broken inversion symmetry in these materials. A ferron is a collective amplitude mode of the spontaneous polarization fluctuations in displacive ferroelectrics [21], including both volume-type [24] and surface-type excitation modes, which carry electric dipoles [25, 26]. Compared to the ferroelectric soft optical phonons, which are quanta of a coherent polarization wave with well-defined phase and frequency [20, 27, 28] and hence also termed "coherent ferrons" [24, 29], ferrons are typically not associated with a coherent wave and therefore they are incoherent. Further on the concept of ferron was extended to the vector fluctuations of the spontaneous polarization considering zero-point oscillations and thermal fluctuations by Yang and Chen [30].

As has been shown recently [31], the influence of the flexocoupling on the fluctuations of electric polarization and elastic strains can lead to the principal changes of the dispersion law of soft optical and acoustic phonons (abbreviated as "*flexophonons*") and the spectral density of ferrons (abbreviated as "*flexoferrons*") in ferroelectrics, using the van der Waals (vdW) ferrielectric $CuInP_2S_6$ as an example. Analytical results, derived in the framework of LGD approach, revealed that the flexophonon and flexoferron dispersion are strongly anisotropic (i.e., depends on the direction of the wavevector $\boldsymbol{k}$) in vdW ferroelectrics and should depend on the magnitude and direction of the external electric field due to the electric-field-induced change in the lattice polarization.

It was shown experimentally by Wooten et al. [32], that an applied electric field changes the velocity of the longitudinal acoustic ferrons, which carry heat and polarization. However, to the best of our knowledge, the influence of electric field on the anisotropic dispersion of the flexophonons and flexoferrons in vdW ferroelectrics has not been studied.

In this work we study theoretically the influence of applied electric field on the anisotropic dispersion of the flexophonons and flexophonons in a uniaxial vdW ferrielectric $CuInP_2S_6$. We consider $CuInP_2S_6$ because the joint action of its multi-well potential energy landscape [33, 34, 35] and negative electrostriction coupling [36] leads to a set of interesting phenomena in this material, such as the anomalous dynamics of polarization in thin films and nanoparticles [37, 38], the temperature and strain tunability of the multiple energy-degenerate metastable polar states [39, 40], and possible emergence of controllable negative capacitance state [41, 42]. The choice of $CuInP_2S_6$ was also motivated by its giant flexoelectric effect, which determines domain engineering [43, 44], mechanical and optoelectronic properties [45, 46] of the ferrielectric.

## 2. Problem Formulation and Basic Equations

Hereinafter we regard that $CuInP_2S_6$ as a uniaxial ferrielectric, whose spontaneous polarization $P_3$ is directed normally to the layers along the $x_3$-axis. The $x_2$ axis is the second-order symmetry axis in the paraelectric phase, which is normal to the monoclinic plane "$m$" in the ferrielectric phase. An external electric field $\vec{E}^{ext}$ is co-directed with the ferrielectric polarization $P_3$. A sketch of $CuInP_2S_6$ layers with crystallographic coordinate system is shown in **Fig. 1(a)**. We would like to note that we oversimplified a real physical picture for $CuInP_2S_6$, because there is an in-plane component of the ferrielectric polarization $P_1$, and the polar axis is close to z-axis, but does not coincide with it.

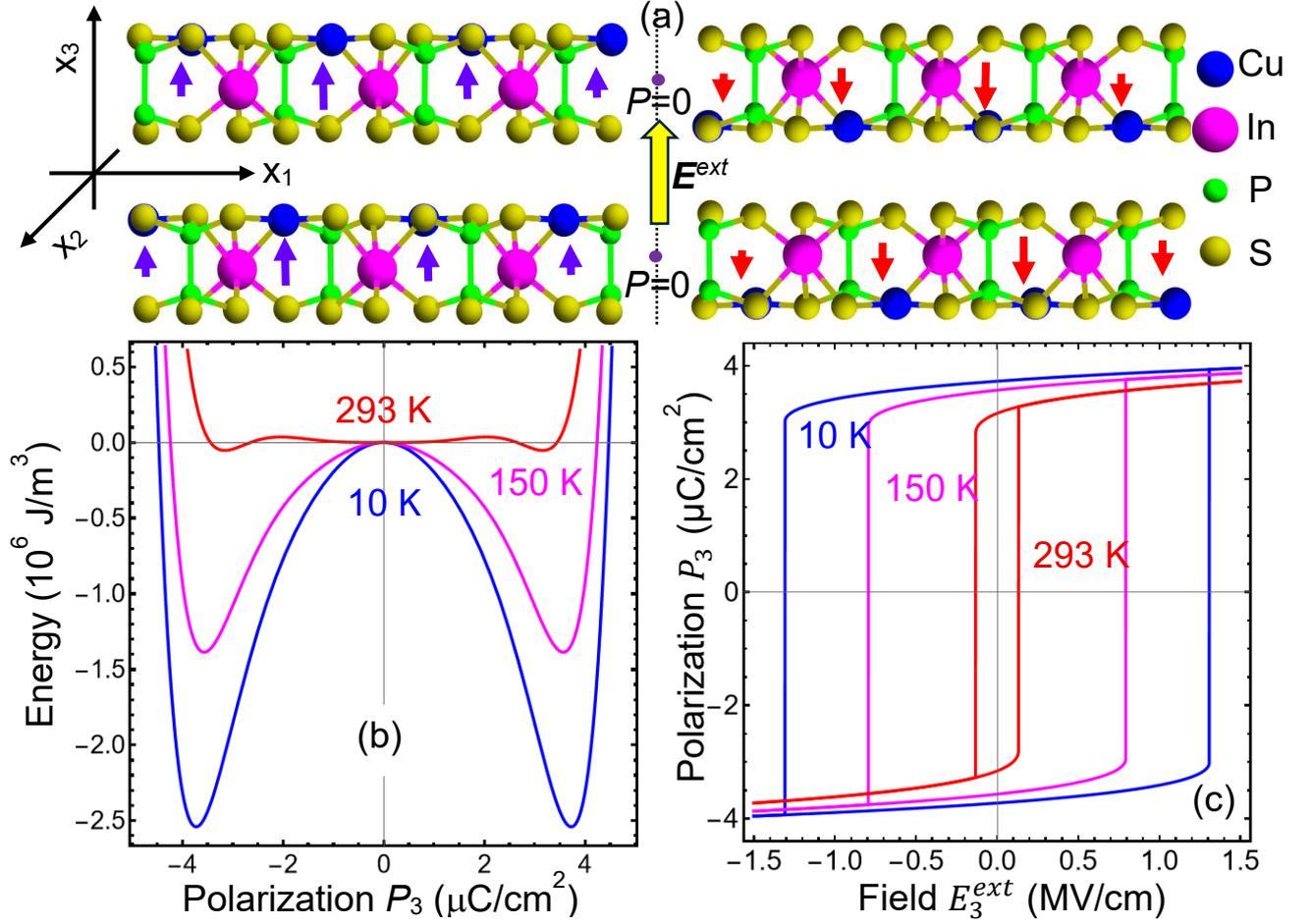

**FIGURE 1. (a)** The sketch of CuInP$_2$S$_6$ layers with crystallographic coordinate system. The axis $x_3$ is normal to the layers, and the axes $x_1$ and $x_2$ are in-plane of the layers. Blue and red arrows with varying length illustrate the transverse fluctuations of the ferrielectric polarization, $\delta P_3(x_1)$, which wavevector **k** is directed along $x_1$-axis. The vertical dotted line shows the break between the counter-polarized layers and corresponds to a virtual plane with zero polarization. **(b)** Dependences of the CuInP$_2$S$_6$ free energy density on the polarization $P_3$ calculated for zero stresses and electric field. The red, magenta and blue curves correspond to different temperatures $T = 293$ K, 150 K and 10 K, respectively. **(c).** Hysteresis loops of the polarization $P_3(E_3^{ext})$ calculated for $T = 10$, 150 and 293 K (blue, magenta and red loops, respectively). Part **(a)** is reproduced from [A. N. Morozovska, E. A. Eliseev, O. V. Bereznikov, M. Ye. Yelisieiev, G.-D. Zhao, Y. Zhu, V. Gopalan, L.-Q. Chen, J.-M. Hu, and Y. M. Vysochanskii, Flexo-coupling induced phonons and ferrons in Van der Waals ferroelectrics, Physical Review B, 112, 014110 (2025)]; licensed under a Creative Commons Attribution (CC BY) license.

Following Ref. [18-19, 31] the Lagrange function of CuInP$_2$S$_6$, $L = \int_t dt \int_{-\infty}^{\infty} dx \, (F - K)$, consists of the kinetic energy $K$ and potential free energy $F$, which are listed in **Appendix A1** [47]. The density of kinetic energy is [18]:

$$K = \frac{\mu}{2}\left(\frac{\partial P_3}{\partial t}\right)^2 + M \frac{\partial P_3}{\partial t}\frac{\partial U_3}{\partial t} + \frac{\rho}{2}\left(\frac{\partial U_3}{\partial t}\right)^2. \qquad (1)$$

Here the coefficient $\mu$ is the polarization inertia [23], $M$ is the magnitude of the dynamic flexocoupling [48] and $\rho$ is the mass density of CuInP$_2$S$_6$. Hereinafter we consider only the coupling between the polarization component $P_3$ and elastic displacement component $U_3$ (which corresponds to the coupling of $P_3$ with the strain components $u_{33}$, $u_{32}$ and $u_{31}$). Following Tagantsev [49], from microscopic point of view, the static bulk ftexoelectricity contribution arises due to the spatial dispersion, or, in other words, due to discontinuity of the crystal lattice. The dynamic bulk flexoelectricity contribution arises due to the frequency dispersion, or in other words, due to a nonequal distribution of mass among the ions of the unit cell.

The free energy $F$ consists of the Landau energy, polarization gradient energy, electrostriction and flexoelectric coupling energies, electrostatic and elastic energy, which are also listed **Appendix A1**. The Landau energy of CuInP$_2$S$_6$ in the 8-th order expansion on the polarization powers, which describes the appearance of the multi-well potential with lower and higher states of the spontaneous polarization in the definite temperature and/or strain range (see e.g., Refs.[37-42] for detail). Using the Gibbs free energy density of the non-polar parent phase as the reference, the free energy density of a single-domain CuInP$_2$S$_6$ under zero external stress can be written as [31]:

$$F(P_3) = \frac{\alpha(T)}{2}P_3^2 + \frac{\beta}{4}P_3^4 + \frac{\gamma}{6}P_3^6 + \frac{\delta}{8}P_3^8 - P_3 E_3^{ext} - \frac{P_3 E_3^d}{2}. \qquad (2)$$

We assume that the temperature dependence of the coefficient $\alpha$ is given by the Barrett-type expression valid from law to high temperatures, $\alpha(T) = \alpha_T T_q \left( \coth \frac{T_q}{T} - \coth \frac{T_q}{T_C} \right)$, where $T_C$ is the Curie temperature and $T_q$ is the quantum vibration temperature. The coefficients $\beta > 0$, $\gamma < 0$ and $\delta > 0$ in Eq.(2) are supposed to be temperature independent. Landau expansion coefficients are listed in **Tables AI-AII** in **Appendix A2** [47]. $E_3^{ext}$ is $x_3$-component of applied electric field co-directed with the ferrielectric polarization $P_3$. $E_3^d$ is z-component of the depolarization field, whose contribution to the Gibbs energy, $\frac{P_3 E_3^d}{2}$, suppresses strongly the longitudinal fluctuations of polarization $\delta P_3(x_3)$ with nonzero derivative $\frac{\partial}{\partial x_3}\delta P_3(x_3)$. Dependences of the Landau energy density, $\frac{\alpha(T)}{2}P_3^2 + \frac{\beta}{4}P_3^4 + \frac{\gamma}{6}P_3^6 + \frac{\delta}{8}P_3^8$, on the polarization $P_3$ calculated for CuInP$_2$S$_6$ parameters and temperatures $T =$ 10, 150 and 293 K are shown in **Fig. 1(b)**. From the figure, the Landau energy has two potential wells at low temperatures, which continuously transform into four wells and then three wells with increase in $T$.

Considering the Tani mechanism and Khalatnikov relaxation, the explicit form of the LGD-KT equations is

$$\Lambda \frac{\partial U_3}{\partial t} = -\frac{\delta L}{\delta U_3}, \qquad \Gamma \frac{\partial P_3}{\partial t} = -\frac{\delta L}{\delta P_3}, \qquad (3)$$

where $\Gamma$ and $\Lambda$ are phenomenological damping constants.

Equations (3) linearized with respect to $P_3$ and $U_3$ allow calculating the generalized susceptibility of CuInP$_2$S$_6$ to the electric field and elastic force (see Eqs. (A.3) in **Appendix A2** [47]). The characteristic equation for the frequency dispersion, $\omega(\mathbf{k})$, of the flexophonons was derived from the singularity of the generalized susceptibility in the Fourier $\{\mathbf{k}, \omega\}$-space (see Ref. [31] and **Appendix A2** [47] for details). For negligibly small or absent damping (i.e., at $\Gamma = \Lambda = 0$), the frequency $\omega(\mathbf{k})$ can be represented in the form [31]:

$$\omega_{O,A}^2(\mathbf{k}) = \frac{1}{2(\mu\rho - M^2)}\left[C(\mathbf{k}) \pm \sqrt{C^2(\mathbf{k}) - 4(\mu\rho - M^2)B(\mathbf{k})}\right]. \tag{4}$$

The signs "+" and "−" before the radical in the dispersion relation (4) correspond to optical (**O**) and acoustic (**A**) phonon modes, respectively. The temperature-dependent and electric field-dependent functions $C(\mathbf{k})$ and $B(\mathbf{k})$ are used:

$$C(\mathbf{k}) = \alpha_0 \rho + (\hat{c}\mathbf{k}^2 \mu - 2\hat{f}\mathbf{k}^2 M + \hat{g}\mathbf{k}^2 \rho) + \mu\hat{v}\mathbf{k}^4, \tag{5a}$$

$$B(\mathbf{k}) = \alpha_0 \hat{c}\mathbf{k}^2 - 4(P_3^0)^2[\hat{q}\mathbf{k} + 2\hat{z}\mathbf{k}(P_3^0)^2]^2 + \hat{c}\mathbf{k}^2 \hat{g}\mathbf{k}^2 + \alpha_0 \hat{v}\mathbf{k}^4 - (\hat{f}\mathbf{k}^2)^2 + \hat{g}\mathbf{k}^2 \hat{v}\mathbf{k}^4. \tag{5b}$$

The temperature-dependent and electric field-dependent positive function $\alpha_0$ is introduced as:

$$\alpha_0 = \alpha^* + 3\beta^*(P_3^0)^2 + 5\gamma(P_3^0)^4 + 7\delta(P_3^0)^6. \tag{6}$$

The coefficients $\alpha^* = \alpha - 2(q_{53}u_{31}^0 + q_{33}u_{33}^0)$ and $\beta^* = \beta - 4(z_{533}u_{31}^0 + z_{333}u_{33}^0)$ are renormalized by elastic strains $u_{3j}^0$. $P_3^0$ is the homogeneous part of the ferrielectric polarization (either spontaneous or field-induced) and $u_{3j}^0$ is the homogeneous part of the strains, which obey the following equations:

$$\alpha^* P_3^0 + \beta^*(P_3^0)^3 + \gamma(P_3^0)^5 + \delta(P_3^0)^7 = E_3^{ext}. \tag{7a}$$

$$c_{55}u_{31}^0 + c_{35}u_{33}^0 = q_{53}(P_3^0)^2 + z_{533}(P_3^0)^4, \quad c_{33}u_{31}^0 + c_{35}u_{31}^0 = q_{33}(P_3^0)^2 + z_{333}(P_3^0)^4. \tag{7b}$$

Hysteresis loops of the polarization $P_3^0(E_3^{ext})$ are shown in **Fig. 1(c)** for the temperatures $T$ from 10 to 293 K. From the figure, the coercive field $E_3^C$ decreases strongly from $\pm 1.31$ MV/cm to $\pm 0.15$ MV/cm with increase in $T$ from 10 K to 293 K. At the same time, the spontaneous polarization $P_3^S$ decreases much weaker with increase in $T$.

The tensorial convolutions $\hat{c}\mathbf{k}^2 = c_{3i3j}k_ik_j$, $\hat{v}\mathbf{k}^4 = v_{3i j3lm}k_ik_jk_lk_m$, $\hat{f}\mathbf{k}^2 = f_{3i3j}k_ik_j$, $\hat{q}\mathbf{k} = g_{3i33}k_i$, $\hat{z}\mathbf{k} = z_{3i33}k_i$ and $\hat{g}\mathbf{k}^2 = g_{3i3j}k_ik_j + \frac{k_3^2}{\varepsilon_0\varepsilon_b k^2}$, are listed in **Appendix A2** [47] for the case of the 2/m symmetry. The coefficients $c_{ijkl}$ are the components of the elastic stiffness tensor; coefficients $v_{ijk}$ are the coefficients of strain gradient energy. Coefficients $f_{ijkl}$ are the components of the static flexocoupling tensor; coefficients $q_{ijkl}$ and $z_{ijkl}$ are the second and higher order electrostriction coupling coefficients, respectively.

Coefficients $g_{ijkl}$ are the polarization energy gradient coefficients. Notably that the last term in $\hat{g}\mathbf{k}^2$, namely $\frac{k_3^2}{\varepsilon_0\varepsilon_b k^2}$, is related to the depolarization field $E_3^d \cong -\frac{\delta P_3}{\varepsilon_0\varepsilon_b}$. Due to the depolarization field $E_3^d$, the longitudinal fluctuations of polarization $\delta P_3(x_3)$ are much smaller than transverse fluctuations $\delta P_3(x_1, x_2)$ [50]. In other words, the $P_3$ is spatially almost uniform along the $x_3$ axis and would remain uniform during the oscillation with consideration of depolarization field, and this is particularly true in atomically layer thick or thin-film CuInP$_2$S$_6$. Thus, we set $k_3 = 0$ for the discussion of the results in the **k**-space. In a particular case $k_3 = 0$ the tensorial convolutions acquire the form:

$$\hat{c}\boldsymbol{k}^2 = c(\varphi)k_\perp^2, \quad \text{where} \quad c(\varphi) = c_{55}\cos^2\varphi + c_{44}\sin^2\varphi, \tag{8a}$$

$$\hat{v}\boldsymbol{k}^4 = v(\varphi)k_\perp^4, \quad \text{where} \quad v(\varphi) = v_{5151}\cos^4\varphi + v_{4242}\sin^4\varphi, \tag{8b}$$

$$\hat{f}\boldsymbol{k}^2 = f(\varphi)k_\perp^2, \quad \text{where} \quad f(\varphi) = f_{55}\cos^2\varphi + f_{44}\sin^2\varphi, \tag{8c}$$

$$\hat{g}\boldsymbol{k}^2 = g(\varphi)k_\perp^2, \quad \text{where} \quad g(\varphi) = g_{55}\cos^2\varphi + g_{44}\sin^2\varphi, \tag{8d}$$

$$\hat{q}\boldsymbol{k} = q_{53}\cos\varphi\, k_\perp, \quad \hat{z}\boldsymbol{k} = z_{533}k_\perp \cos\varphi, \tag{8e}$$

where $k^2 = k_1^2 + k_2^2 \equiv k_\perp^2$ and $\tan\varphi = \frac{k_1}{k_2}$.

Like optical and acoustic flexophonons, the flexoferrons can be classified into acoustic and optical flexoferrons [31]. Following Ref. [23], we consider the case when the mechanical force and the electric field in Eqs.(3) are Langevin-type noise fields, which obey the fluctuation-dissipation theorem for fluctuation dynamics [51] (see also **Appendix A3** [47] for details). Assuming that the damping constant is small and using the perturbation theory, in Ref. [31] we derived the second-order correction for the polarization response to the Langevin-type electric noise:

$$\langle p \rangle \approx \int_{-\infty}^{\infty} \frac{d^3k}{(2\pi)^3} \left[ \coth\left(\frac{\hbar\omega_O(\boldsymbol{k})}{2k_B T}\right) \delta p_O(\boldsymbol{k}) + \coth\left(\frac{\hbar\omega_A(\boldsymbol{k})}{2k_B T}\right) \delta p_A(\boldsymbol{k}) \right], \tag{9}$$

where $\delta p_O(\boldsymbol{k})$ and $\delta p_A(\boldsymbol{k})$ are the spectral densities of optical and acoustic flexoferrons, the eigen frequencies $\omega_O(\boldsymbol{k})$ and $\omega_A(\boldsymbol{k})$ are given by Eq.(4). Approximate analytical expressions for $\delta p_O(\boldsymbol{k})$ and $\delta p_A(\boldsymbol{k})$ are [31]:

$$\delta p_O(\boldsymbol{k}) \approx \frac{-\hbar}{2(\mu\rho - M^2)} \frac{\eta}{\alpha_0} \left| \frac{\hat{v}\boldsymbol{k}^4 + \hat{c}\boldsymbol{k}^2 - \rho\omega_O^2(\boldsymbol{k})}{\left(\omega_A^2(\boldsymbol{k}) - \omega_O^2(\boldsymbol{k})\right)\omega_O(\boldsymbol{k})} \right|, \tag{10a}$$

$$\delta p_A(\boldsymbol{k}) \approx \begin{cases} \frac{-\hbar}{2(\mu\rho - M^2)} \frac{\eta}{\alpha_0} \left| \frac{\hat{v}\boldsymbol{k}^4 + \hat{c}\boldsymbol{k}^2 - \rho\omega_A^2(\boldsymbol{k})}{\left(\omega_O^2(\boldsymbol{k}) - \omega_A^2(\boldsymbol{k})\right)\omega_A(\boldsymbol{k})} \right|, & \omega_A^2(\boldsymbol{k}) > 0, \\ \frac{-\hbar}{2(\mu\rho - M^2)} \frac{\eta}{\alpha_0} \left| \frac{\hat{v}\boldsymbol{k}^4 + \hat{c}\boldsymbol{k}^2 - \rho\omega_A^2(\boldsymbol{k})}{\left(\omega_O^2(\boldsymbol{k}) - \omega_A^2(\boldsymbol{k})\right)\omega_A(\boldsymbol{k})} \right| \frac{\Gamma}{\sqrt{\Gamma^2 - 4\mu^2\omega_A^2(\boldsymbol{k})}}, & \omega_A^2(\boldsymbol{k}) < 0. \end{cases} \tag{10b}$$

Here the field-dependent function $\eta(E_3^{ext}) = 3\beta^* P_3^0 + 10\gamma(P_3^0)^3 + 21\delta(P_3^0)^5$ is introduced. Note that $\omega_O^2(\boldsymbol{k})$ is always positive, but the case $\omega_A^2(\boldsymbol{k}) < 0$ is possible [31] for high flexoelectric coefficients and/or high electric fields close to the coercive field. From the second line of Eq.(10b), valid for $\omega_A^2(\boldsymbol{k}) < 0$, we obtain that $\delta p_A(\boldsymbol{k}) \to 0$ at $\Gamma \to 0$, being consistent with impossibility to quantize the elementary oscillator with imaginary frequency $\omega_A(\boldsymbol{k})$ due to its damping.

Since we are interested in flexoferron properties in the external field, below we discuss results for low temperatures ($T \leq 10$ K). Due to the electric noise source, ferrons considered in this work (as well as in Ref. [31]) are incoherent.

## 3. Results and Discussion

The optical flexophonon frequency $\omega_O(\mathbf{k})$ is always positive (see e.g., the right columns in **Figs. B1** and **B3** in **Appendix B** [47]). The acoustic flexophonon frequency $\omega_A(\mathbf{k})$ becomes purely imaginary for definite nonzero values of wavevectors, flexoelectric coefficients and/or electric field (see e.g., the left columns in **Figs. B1** and **B3** in **Appendix B** [47]).

The threshold condition $\omega_A(\mathbf{k}) = 0$ at nonzero $\mathbf{k}$ may correspond to the transition into incommensurate spatially modulated phase induced by the flexocoupling [19, 31]. From Eq.(4) the condition for the determination of the field-dependent critical wavevector $\mathbf{k}_{cr}$ and flexoelectric coefficient $f_{cr}(\mathbf{k})$ is the equation $B(\mathbf{k}) = 0$, which explicit form for $k_\perp^2 \neq 0$ and $k_3 = 0$ is

$$g(\varphi)v(\varphi)k_\perp^4 + \left(v(\varphi)\alpha_S - f(\varphi)^2 + c(\varphi)g(\varphi)\right)k_\perp^2 + c(\varphi)\alpha_0 - 4(P_3^0)^2(q_{53} + 2(P_3^0)^2 z_{533})^2\cos^2\varphi = 0. \quad (11)$$

The dependence of the flexoelectric coefficient $f$ on the wavevector $\mathbf{k}$ corresponding to the condition $\text{Re}[\omega_A(\mathbf{k})] = 0$, are shown in **Fig. 2(a) – 2(c)**, respectively. Color plots are calculated for $T = 10$ K and several relatively high values of external field (close to coercive field $E_3^c = -1.31$ MV/cm). The red color in the plots corresponds to the cropping of the color scale at the level $f = 10$ V. Note that $f \gg 10$ V inside the red regions and diverges ($f \to \infty$) approaching their central parts. From **Fig. 2(a)** calculated for $E_3^{ext} = -1$ MV/cm, the condition $\text{Re}[\omega_A(\mathbf{k})] = 0$ is possible only for relatively high $f > 5$ V. The physical picture at lower $E_3^{ext}$ looks like the one shown in **Fig. 2(a)**. From **Fig. 2(b)** calculated for $E_3^{ext} = -1.15$ MV/cm, the condition $\text{Re}[\omega_A(\mathbf{k})] = 0$ is possible for much smaller $f$, and very thin dark-violet stripes, where $f \to 0$, appear at small $k_2 \to 0$ and $0.1$ nm $< k_1 < 0.2$ nm. The stripes become thicker and longer and eventually transform into a bow-like region with further increase in $E_3^{ext}$. A large bow-like dark-violet region, where $f = 0$, corresponds to $E_3^{ext} = -1.3$ MV/cm (see **Fig. 2(c)**). Hence $\mathbf{k}$-region, where $\text{Re}[\omega_A(\mathbf{k})] = 0$ at $f = 0$, appears and expands with increase in electric field near the coercive value. Simultaneously, the red region, where $f \gg 10$ V, shrinks strongly.

Using biquadratic Eq.(11) we derived the analytical expression for the field-dependent critical value $k_{cr}$ in **Appendix A2** [47], which is

$$k_{cr}^2(\varphi) = \frac{1}{2g(\varphi)v(\varphi)}\left[f^2(\varphi) - f_{th}^2(\varphi) \pm \sqrt{[f^2(\varphi) - f_{th}^2(\varphi)]^2 - \xi_{sm}(\varphi)}\right], \quad (12a)$$

where the value $f_{th}(\varphi)$ is an upper thermodynamic limit [11, 19] of the flexoelectric coefficient

$$f_{th}^2(\varphi) = v(\varphi)\alpha_0 + c(\varphi)g(\varphi). \quad (12b)$$

Indeed, the condition $v(\varphi) = 0$ leads to the expression $f_{cr}^2(\varphi) = c(\varphi)g(\varphi)$, that agrees with the maximal values of the static flexoelectric coefficient (the "upper thermodynamic limit") established in Ref. [10].

The value $\xi_{sm}(\varphi)$ is introduced as

$$\xi_{sm}(\varphi) = 4g(\varphi)v(\varphi)[c(\varphi)\alpha_0 - 4(P_3^0)^2(q_{53} + 2(P_3^0)^2 z_{533})^2 \cos^2\varphi]. \quad (12c)$$

The critical value of the flexocoupling coefficient is

$$f_{cr}^2(\varphi) = \begin{cases} f_{th}^2(\varphi) + \sqrt{\xi_{sm}(\varphi)}, & \xi_{sm}(\varphi) \geq 0, \\ 0, & \xi_{sm}(\varphi) < 0. \end{cases} \quad (13)$$

Below we consider the case

$$f_{44} \approx f_{55} \cong f \quad \text{and} \quad g_{44} \approx g_{55} \cong g, \quad (14)$$

which is consistent with CuInP$_2$S$_6$ symmetry. In this case $f(\varphi)$ and $g(\varphi)$ becomes angle-independent according to their definitions, Eqs.(8c) and (8d).

The critical value of the flexoelectric coefficient $f_{cr}$ as a function of the external field $E_3^{ext}$ and the orientation angle $\varphi$ of wavevector $\boldsymbol{k}$ is shown in **Fig. 2(d)**. The value of $f_{cr}$ is maximal at $E_3^{ext} = 0$ and decreases monotonically with increase in $E_3^{ext}$ up to the coercive value $E_3^c \approx -1.31$ MV/cm. The dark-violet region, where $f_{cr} = 0$ due to $\xi_{sm}(\varphi) < 0$, appears at $E_3^{ext} > E_3^{th}$, where the threshold value $E_3^{th} \approx -1.1$ MV/cm. The boundary of the region, where $f_{cr} = 0$, corresponds to the condition $c(\varphi)\alpha_0 - 4(P_3^0)^2[q_{53} + 2(P_3^0)^2 z_{533}]^2 \cos^2\varphi = 0$. The value of $f_{cr}$ is maximal for the angles $\varphi = \frac{\pi n}{2}$ and minimal for $\varphi = \frac{\pi}{2} + \frac{\pi n}{2}$ ($n$ is an integer number). The region, where $f_{cr} = 0$, exists at the angles $-\frac{3\pi}{8} \leq \varphi \leq \frac{3\pi}{8}$.

The "gap" in the dependence $f_{cr}(\varphi)$, where $f_{cr} = 0$, is clearly seen at the two lowest curves in **Fig. 2(e)**, calculated for $E_3^{ext} = -1.2$ MV/cm and $E_3^{ext} = -1.3$ MV/cm, respectively. The angular dependence $f_{cr}(\varphi)$ is quasi-sinusoidal for $E_3^{ext} < E_3^{th}$, namely $f_{cr}(\varphi) \approx f_{cr}^0 + \delta f \cdot \cos(\varphi)$, where the "base" $f_{cr}^0$ and the amplitude $\delta f$ increase with decrease in $E_3^{ext}$. The angular anisotropy of $f_{cr}(\varphi)$ is clearly seen from the polar plot **Fig. 2(f).** The principal change of the contours shape appears at $E_3^{ext} < E_3^{th}$.

It is seen from **Fig. 2**, that the critical flexoelectric coefficient $f_{cr}(\varphi)$ is largely determined by the magnitude of $E_3^{ext}$ and by the wavevector direction expressed by $\cos(\varphi)$. This allows us to propose a method for estimating the strength of the flexocoupling in ferrielectric materials, providing that the frequency of acoustic flexophonons is zeroing at the threshold value $E_3^{th}$ of applied electric field. The suggestion agrees with the recent finding in SrTiO$_3$ [20].

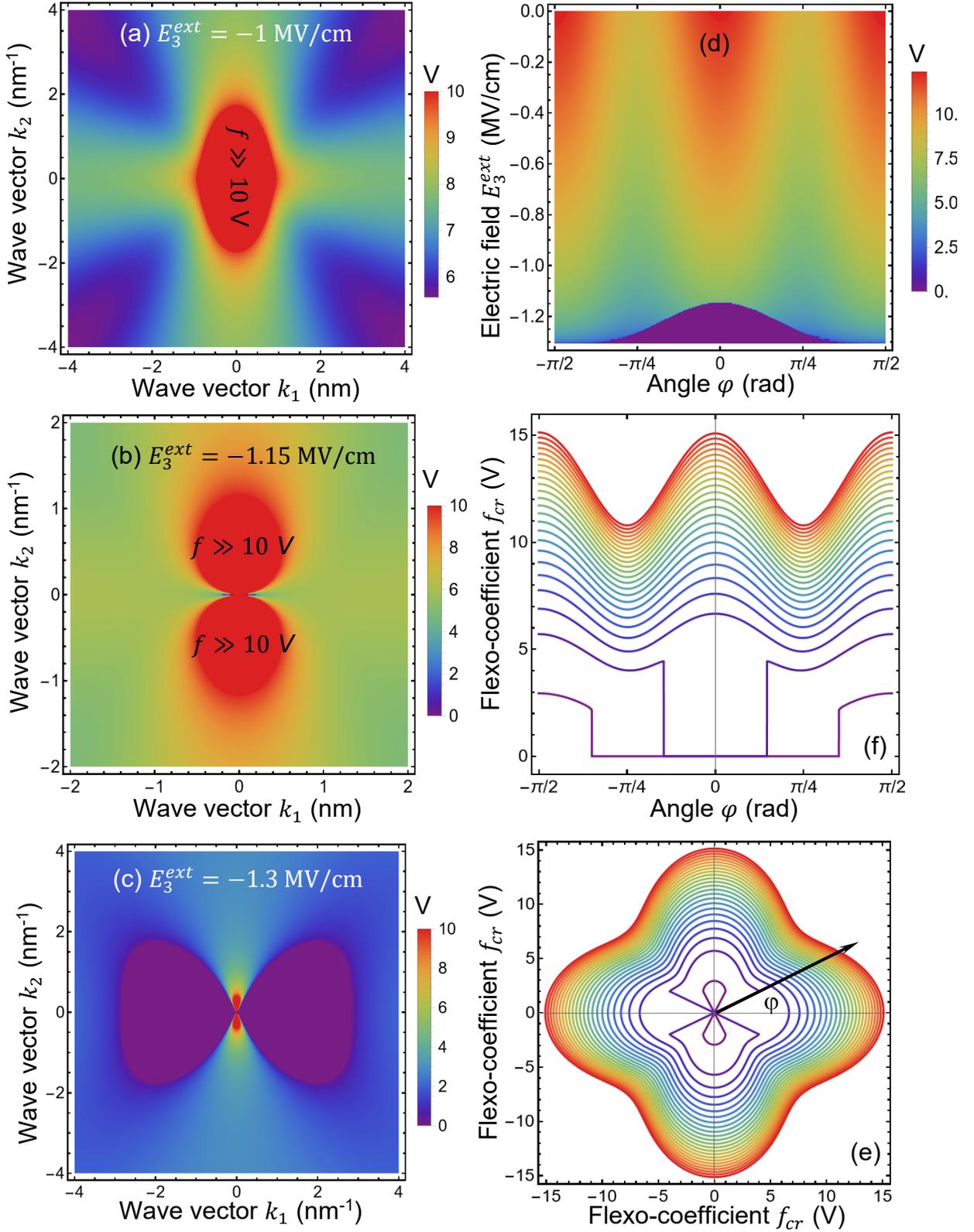

**FIGURE 2**. The dependence of the flexoelectric coefficient $f$ on the wavevector $\mathbf{k}$ corresponding to the condition $Re[\omega_A] = 0$, calculated for several values of applied field $E_3^{ext}$ =-1 MV/cm **(a)**, -1.15 MV/cm **(b)** and -1.3 MV/cm **(c)**. **(d)** The critical value of the flexoelectric coefficient $f_{cr}$ as a function of $E_3^{ext}$ and $\varphi$. **(e)** The dependence of $f_{cr}$ and as a function of $\varphi$ calculated for $E_3^{ext}$ changing from -1.3 MV/cm to +1 MV/cm with the step of 0.1 MV/cm (from violet to red curves). **(f)** The polar plot of $f_{cr}(\varphi)$ calculated for $E_3^{ext}$ changing from -

1.3 MV/cm to +1 MV/cm with the step of 0.1 MV/cm (from violet to red contours). The damping is absent ($\Gamma = 0, \Lambda = 0$), $k_3 = 0$ and $T = 10$ K for all plots. Material parameters of CuInP$_2$S$_6$ are listed in **Table AI** [47].

It should be noted that the symmetry properties of CuInP$_2$S$_6$ are similar to the monoclinic uniaxial ferroelectric Sn$_2$P$_2$Se$_6$ [52], where the spatially modulated incommensurate phase exists, but narrows its temperature range with increase in the electric field strength directed along the spontaneous polarization and disappears above a certain critical strength of the field. Complementary, it may seem that an applied electric field can induce a spatially modulated phase in CuInP$_2$S$_6$. However, based on the above results, we would like to underline that only the fields very close to the coercive one (but slightly less than it) can induce the spatially modulated phase. Fields, which are significantly greater or significantly less than the coercive field, do not make any positive contribution to the modulated phase appearance. Thus, one needs to lower the polarization with the field, and where it should abruptly change the sign (i.e., at the coercive field), the acoustic phonon frequency tends to zero and the spatial modulation may occur simultaneously. Perhaps the electric field changes the energy balance between ferrielectric and antiferroelectric order parameters [53], which determine the polar properties of CuInP$_2$S$_6$ [35, 37], and thus it changes the local potential and the frequencies of optical phonons in the center of the Brillouin zone and the speed of sound [52, 54]. Hence, a dynamic instability near the center of the Brillouin zone may arise.

The real part of the acoustic phonon frequency, Re[$\omega_A$], as the function of the wavevector components $k_1$ and $k_2$, is shown in **Fig. 3**. The color plots are calculated for T = 10 K, several values of the flexoelectric coefficient $f$, which changes from 0 to 10 V, and electric field $E_3^{ext}$, which changes from 0 MV/cm to 1.2 MV/cm. **Figures B1** and **B3** in **Appendix B** [47] show the *k*-behavior of the frequency $\omega_O$, that depends weakly on the flexocoupling and external field, for comparison. The frequency $\omega_A$ is an angular-dependent positive function of ***k*** for $f < f_{cr}(E_3^{ext})$. The magnitude of $\omega_A$ increases with increase in ***k*** for $f < f_{cr}$ (corresponding regions are shown by the rainbow color scale in **Fig. 3**). The frequency $\omega_A$ becomes zero at $f = f_{cr}(E_3^{ext})$ and purely imaginary at $f > f_{cr}(E_3^{ext})$ (corresponding regions are shown by the dark-violet color in **Fig. 3**). The regions, where Re[$\omega_A$] = 0 at nonzero ***k***, are absent when the flexoelectric coefficients are small (e.g., less than 8 V) and electric fields are well below the coercive one. The increase of electric field and/or the increase of $f$ changes the shape of the dark-violet regions, where Re[$\omega_A$] = 0, from the horizontal bow-shaped region (as shown in **Fig. 3(c)** and **3(f)**) to the cross-shaped region (**Fig. 3(j)**), then to the whole ***k***-plane without a large central oval region (**Fig. 3(h)** and **3(k)**), and eventually to the ***k***-plane without two vertically aligned small ovals (see e.g., **Fig. 3(i)** and **3(l)**).

The spectral density of acoustic flexoferrons, $\delta p_A$, as the function of the wavevector components $k_1$ and $k_2$, is shown in **Fig. 4**. Color plots are calculated for T = 10 K, several values of the flexoelectric coefficient $f$, which changes from 0 to 10 V, and electric field $E_3^{ext}$, which changes from 0 MV/cm to 1.2 MV/cm. **Figures B2** and **B4** in **Appendix B** [47] show the $\boldsymbol{k}$-behavior of the spectral density of optic flexoferrons, $\delta p_0$, that depends very weakly on the flexoelectric coupling and external field in comparison with $\delta p_A$. The density $\delta p_A$ is a weakly angular-dependent positive function of $\boldsymbol{k}$ for $f < f_{cr}(E_3^{ext})$. The magnitude of $\delta p_A$ increases strongly with increase in $\boldsymbol{k}$ for $f < f_{cr}$ (corresponding regions are shown by the orange and yellowish colors in **Fig. 4**). The density $\delta p_A$ diverges at $f = f_{cr}(E_3^{ext})$ and then becomes zero at $f > f_{cr}(E_3^{ext})$ in accordance with Eq.(13). The dark-violet color in **Fig. 4** corresponds to the cropping of the scale at the level $\delta p_A = -50 \cdot 10^{-32}$ C·m. Thus $\delta p_A \ll -50 \cdot 10^{-32}$ C·m inside the dark-violet regions and diverges approaching the red boundary. The regions, where $\delta p_A = 0$, are shown by the red color. These regions are absent when the flexoelectric coefficients are small (e.g., less than 8 V) and electric fields are well below the coercive ones. The increase of electric field and/or the increase of $f$ changes the shape of the red regions, where $\delta p_A = 0$ at nonzero $\boldsymbol{k}$, from the horizontal bow-shaped region (as shown in **Fig. 4(c)** and **4(f)**) to the cross-shaped region (**Fig. 4(j)**), then to the whole $\boldsymbol{k}$-plane without a large central oval region (**Fig. 4(h)** and **4(k)**), and eventually to the $\boldsymbol{k}$-plane without two vertically aligned small ovals (see e.g., **Fig. 4(i)** and **4(l)**). These changes are determined by the changes of the region, where $\text{Re}[\omega_A] = 0$, being fully consistent with those shown in **Fig. 3**.

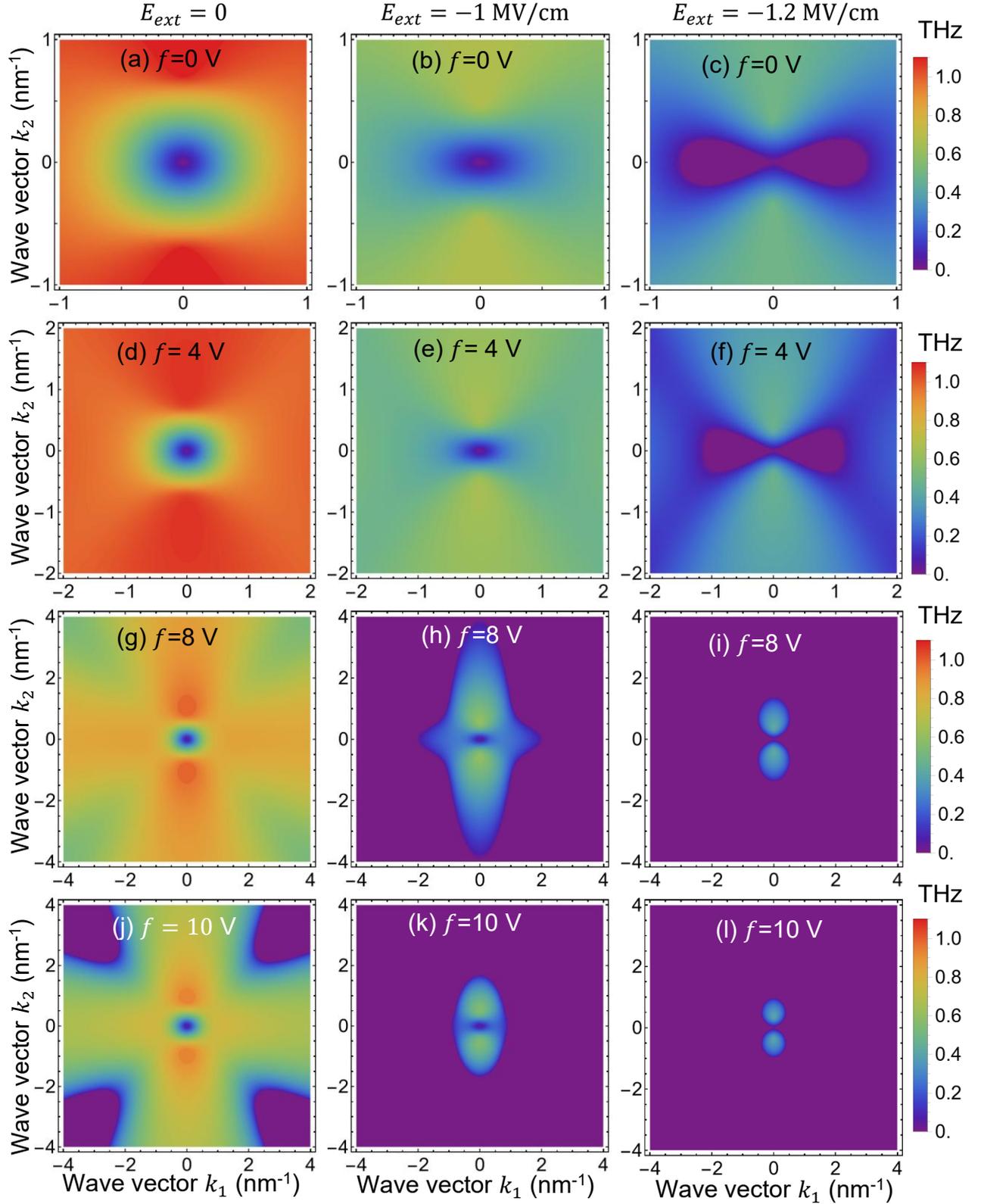

**FIGURE 3.** The real part of the acoustic phonon frequency, $Re[\omega_A]$, as the function of the wavevector components $k_1$ and $k_2$, calculated for several values of the flexoelectric coefficient $f=0$ **(a, b, c)**, 4 **(d, e, f)**, 8 **(g, h, i)** and 10 V **(j, k, l)**. The magnitude of electric field is 0 MV/cm **(a, d, g, j)**, -1 MV/cm **(b, e, h, k)** and -1.2 MV/cm **(c, f, i, l)**. The damping is absent ($\Gamma = 0, \Lambda = 0$), $k_3 = 0$ and $T = 10$ K for all plots. Material parameters of CuInP$_2$S$_6$ are listed in **Table AI** [47].

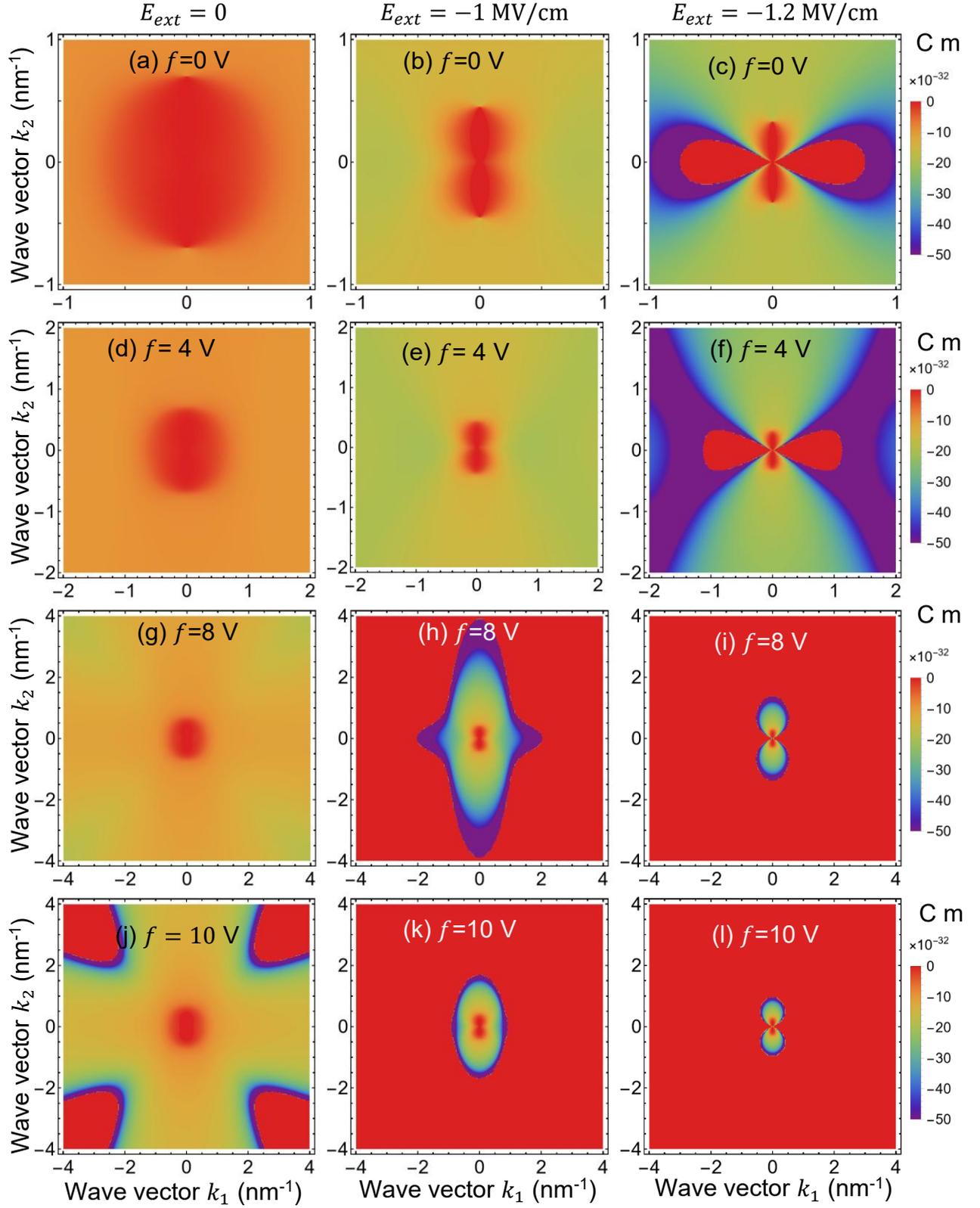

**FIGURE 4.** The spectral density of the acoustic flexoferrons, $\delta p_A$, as the function of the wavevector components $k_1$ and $k_2$, calculated for several values of the flexoelectric coefficient $f_{55}$=0 **(a, b, c)**, 4 **(d, e, f)**, 8 **(g, h, i)** and 10 V **(j, k, l)**. The magnitude of electric field is 0 MV/cm **(a, d, g, j)**, -1 MV/cm **(b, e, h, k)** and -1.2 MV/cm **(c, f, i, l)**. The damping is absent ($\Gamma = 0, \Lambda = 0$), $k_3 = 0$ and $T = 10$ K for all plots. Material parameters of CuInP$_2$S$_6$ are listed in **Table AI** [47].

## 4. Conclusions

In this work we study the influence of applied electric field on the angular-dependent anisotropic dispersion of the flexophonons and flexophonons in a uniaxial vdW ferrielectric $CuInP_2S_6$. We reveal the principal changes in the dispersion law of acoustic flexophonons and flexoferrons at low temperatures, which emerge with increase of applied electric field. In particular, the frequency of acoustic flexophonons and flexoferrons becomes zero at nonzero wavevectors when the field reaches the critical value, and purely imaginary when the field exceeds the value. Corresponding spectral density of acoustic flexoferrons diverges at the critical field and disappears at higher fields.

We relate these changes with a possible appearance of a spatially modulated incommensurate polar phase induced by the increase of the flexocoupling strength. The magnitude of the critical flexoelectric coefficient is determined by the magnitude of the electric field and the direction of the wave vector. It is important that the critical value of the flexocoupling strength becomes zero when external field exceeds the threshold value. This allows us to propose a method for estimating the strength of the flexocoupling in the vdW ferrielectrics from the condition of zeroing the frequency of the acoustic flexophonon, accompanied by the divergency of the flexoferron spectral density, at the threshold value of applied electric field. Since the values of the flexoelectric coefficients are poorly known in vdW ferrielectrics, obtained analytical results can be useful for their flexo-engineering.

**Supplementary Material.** Supporting Information containing calculation details (Appendix A), icluding free energy functional (Appendix A1), analytical solution for the eigen frequency of the flexophonons (Appendix A2) and flexoferrons (Appendix A3), and additional figures (Appendix B) are given in Supplementary Materials.

**Authors' contribution.** A.N.M. generated the research idea, formulated the problem, performed most analytical calculations and wrote the manuscript draft jointly with J-M.H. E.A.E. wrote the codes and prepared figures. All authors worked on the analysis of results and manuscript improvement.

**Acknowledgements.** The work of A.N.M., E.A.E., Y.Z., V.G., L.-Q.C., and J-M.H. is supported by the DOE Software Project on "Computational Mesoscale Science and Open Software for Quantum Materials", under Award Number DE-SC0020145 as part of the Computational Materials Sciences Program of US Department of Energy, Office of Science, Basic Energy Sciences. Y.M.V. acknowledges support from the Horizon Europe Framework Programme (HORIZON-TMA-MSCA-SE), project № 101131229, Piezoelectricity in 2D-materials: materials, modeling, and applications (PIEZO 2D). The work of A.N.M. is also sponsored by the Target Program of the National Academy of Sciences of Ukraine (Project No. 5.8/25-П "Energy-saving and environmentally friendly nanoscale ferroics for the development of sensorics, nanoelectronics and spintronics" and "B/222. Research of optical, polar and electrophysical properties of organic and inorganic hybrid nanostructures and nanomaterials of functional purpose"). Results were visualized in Mathematica 14.0 [55].

## Supplementary Materials
### Appendix A. Calculation Details
### A1. The Free Energy Functional

Ferrielectric and paraelectric phases of CuInP$_2$S$_6$ have the point symmetry $m$ and $2/m$, respectively. Here we use the following coordinate system: the coordinate axis "$x_2$" is along the symmetry axis "2", the coordinate axis "$x_3$" is along the normal vector of all the layers, the coordinate axis "$x_1$" is perpendicular to "$x_3$" and $x_2$". The spontaneous polarization $P_3$ is directed normally to the layers.

Using the scalar approximation in the considered one-component 1D case, the Lagrange function $L = \int_t dt \int_{-\infty}^{\infty} dx \, (F - K)$ consists of the kinetic energy K and free energy $F$ of ferroelectric. The density of kinetic energy is [18]:

$$K = \frac{\mu}{2}\left(\frac{\partial P_3}{\partial t}\right)^2 + M\frac{\partial P_3}{\partial t}\frac{\partial U_3}{\partial t} + \frac{\rho}{2}\left(\frac{\partial U_3}{\partial t}\right)^2. \tag{A.1}$$

Here the coefficient μ is the polarization inertia [23], $M$ is the magnitude of the dynamic flexocoupling [48] and ρ is the mass density of CuInP$_2$S$_6$. Hereinafter we consider only the coupling between the polarization component $P_3$ and elastic displacement component $U_3$.

The bulk density of the free energy $F$ that depends on the polarization $P_3$ and strain components $u_3$, $u_4$ and $u_5$, and their gradients, has the following form in Voigt notations [31]:

$$F = \frac{\alpha}{2}P_3^2 + \frac{\beta}{4}P_3^4 + \frac{\gamma}{6}P_3^6 + \frac{\delta}{8}P_3^8 + \frac{g_{55}}{2}\left(\frac{\partial P_3}{\partial x_1}\right)^2 + \frac{g_{44}}{2}\left(\frac{\partial P_3}{\partial x_2}\right)^2 + g_{35}\frac{\partial P_3}{\partial x_3}\frac{\partial P_3}{\partial x_1} + \frac{g_{33}}{2}\left(\frac{\partial P_3}{\partial x_3}\right)^2 - P_3 E_3^{ext} - \frac{P_3 E_3^d}{2} +$$

$$- q_{53}u_5 P_3^2 - q_{33}u_3 P_3^2 - z_{533}u_5 P_3^4 - z_{333}u_3 P_3^4 +$$

$$+ f_{55}u_5\frac{\partial P_3}{\partial x_1} + f_{53}u_5\frac{\partial P_3}{\partial x_3} + f_{44}u_4\frac{\partial P_3}{\partial x_2} + f_{33}u_3\frac{\partial P_3}{\partial x_3} + f_{35}u_3\frac{\partial P_3}{\partial x_1} +$$

$$+ \frac{c_{55}}{2}u_5^2 + \frac{c_{44}}{2}u_4^2 + c_{35}u_5 u_3 + \frac{c_{33}}{2}u_3^2 + \frac{v_{5151}}{2}\left(\frac{\partial u_5}{\partial x_1}\right)^2 + \frac{v_{4242}}{2}\left(\frac{\partial u_4}{\partial x_2}\right)^2 + \frac{v_{3333}}{2}\left(\frac{\partial u_3}{\partial x_3}\right)^2 - N_3 U_3 \tag{A.2}$$

The gradient coefficients $g_{ij}$ determines the magnitude of the gradient energy. $E_3^{ext}$ is z-component of external electric field co-directed with the polar axis $x_3$. $E_3^d$ is z-component of to the depolarization field, which contribution to the free energy, $\frac{P_3 E_3^d}{2}$, suppresses strongly the longitudinal fluctuations of polarization. The coefficients $q_{ij}$ are second-order electrostriction coupling coefficients, and $z_{ijk}$ are higher-order electrostriction coupling coefficients. Coefficients $f_{ij}$ are the components of the static flexocoupling tensor. Coefficients $c_{ii}$ are elastic stiffness. $N_3$ is z-component of the external mechanical force bulk density.

**A2. Analytical Solution for the Eigen Frequency of the Optical and Acoustic Flexophonons**

Analytical solution for the eigen frequency of the optical and acoustic flexophonons are derived in Ref. [31]. Basic equations and analytical expressions are reproduced below for the sake of the readers' convenience.

Considering the Tani mechanism and Khalatnikov relaxation, the explicit form of the LGD-KT equations, $\delta L/\delta U_3 = 0$ and $\delta L/\delta P_3 = 0$, is [31]:

$$v_{3ij3kl}\frac{\partial^4 U_3}{\partial x_i \partial x_j \partial x_k \partial x_l} + \rho\frac{\partial^2 U_3}{\partial t^2} + \Lambda\frac{\partial U_3}{\partial t} - c_{55}\frac{\partial^2 U_3}{\partial x_1^2} - c_{44}\frac{\partial^2 U_3}{\partial x_2^2} - 2c_{35}\frac{\partial^2 U_3}{\partial x_1 \partial x_3} - c_{33}\frac{\partial^2 U_3}{\partial x_3^2} - f_{55}\frac{\partial^2 P_3}{\partial x_1^2} -$$
$$f_{44}\frac{\partial^2 P_3}{\partial x_2^2} - (f_{35} + f_{53})\frac{\partial^2 P_3}{\partial x_1 \partial x_3} - f_{33}\frac{\partial^2 P_3}{\partial x_3^2} + 2P_3\left(q_{53}\frac{\partial P_3}{\partial x_1} + q_{33}\frac{\partial P_3}{\partial x_3}\right) + 4\left(z_{533}\frac{\partial P_3}{\partial x_1} + z_{333}\frac{\partial P_3}{\partial x_3}\right)P_3^3 +$$
$$M\frac{\partial^2 P_3}{\partial t^2} = N. \quad (A.3a)$$

$$\mu\frac{\partial^2 P_3}{\partial t^2} + \Gamma\frac{\partial P_3}{\partial t} + \alpha P_3 + \beta P_3^3 + \gamma P_3^5 + \delta P_3^7 - g_{55}\frac{\partial^2 P_3}{\partial x_1^2} - g_{44}\frac{\partial^2 P_3}{\partial x_2^2} - 2g_{35}\frac{\partial^2 P_3}{\partial x_1 \partial x_3} - g_{33}\frac{\partial^2 P_3}{\partial x_3^2} - f_{55}\frac{\partial^2 U_3}{\partial x_1^2} -$$
$$f_{44}\frac{\partial^2 U_3}{\partial x_2^2} - (f_{35} + f_{53})\frac{\partial^2 U_3}{\partial x_1 \partial x_3} - f_{33}\frac{\partial^2 U_3}{\partial x_3^2} - 2\left(q_{53}\frac{\partial U_3}{\partial x_1} + q_{33}\frac{\partial U_3}{\partial x_3}\right)P_3 - 4\left(z_{533}\frac{\partial U_3}{\partial x_1} + z_{333}\frac{\partial U_3}{\partial x_3}\right)P_3^3 +$$
$$M\frac{\partial^2 U_3}{\partial t^2} = E_3^{ext} + E_3^d. \quad (A.3b)$$

Here $\Gamma$ and $\Lambda$ are damping constants of polarization and displacement, respectively.

The Fourier integral expansions are used for polarization $P$, displacement $U$, perturbation electric field $E$ and mechanical force density $N$:

$$P = P_3^0 + \int d\omega \int d^3\mathbf{k}\, exp(i\mathbf{k}\mathbf{x} - i\omega t)\, \tilde{P}, \quad U = u_{3j}^0 x_j + \int d\omega \int d^3\mathbf{k}\, exp(i\mathbf{k}\mathbf{x} - i\omega t)\, \tilde{U}, \quad (A.4a)$$

$$E = \int d\omega \int d^3\mathbf{k}\, exp(i\mathbf{k}\mathbf{x} - i\omega t)\, \tilde{E}, \quad N = \int d\omega \int d^3\mathbf{k}\, exp(i\mathbf{k}\mathbf{x} - i\omega t)\, \tilde{N}. \quad (A.4b)$$

Hereinafter $P_3^0$ is the homogeneous part of the ferrielectric polarization (either spontaneous or field-induced) and $u_{3j}^0$ is the homogeneous part of corresponding strain component. For the particular case $\delta = 0, \gamma > 0$ the analytical expressions for the spontaneous polarization and strain exist, namely $P_S^2 = (\sqrt{\beta^2 - 4\alpha\gamma} - \beta)/2\gamma$ and $u_{3j}^0 = \frac{1}{c}(qP_S^2 + zP_S^4)$. In the considered case $P_3^0$ and nonzero $u_{3j}^0$ obey the following equations:

$$[\alpha - 2(q_{53}u_{31}^0 + q_{33}u_{33}^0)]P_3^0 + [\beta - 4(z_{533}u_{31}^0 + z_{333}u_{33}^0)](P_3^0)^3 + \gamma(P_3^0)^5 + \delta(P_3^0)^7 = E_3^{ext}.$$
$$(A.5a)$$

$$c_{55}u_{31}^0 + c_{35}u_{33}^0 = q_{53}(P_3^0)^2 + z_{533}(P_3^0)^4, \quad c_{33}u_{31}^0 + c_{35}u_{31}^0 = q_{33}(P_3^0)^2 + z_{333}(P_3^0)^4. \quad (A.5b)$$

The Fourier representation of linearized Eqs.(A.3) has the form [31]:

$$\left(\hat{v}\mathbf{k}^4 + \hat{c}\mathbf{k}^2 - i\Lambda\omega - \rho\omega^2\right)\tilde{U} + \left(\hat{f}\mathbf{k}^2 + 2i\hat{q}\mathbf{k}P_S + 4i\hat{z}\mathbf{k}P_S^3 - M\omega^2\right)\tilde{P} = \tilde{N}, \quad (A.6a)$$

$$\left(-i\Gamma\omega - \mu\omega^2 + \alpha_0 + \hat{g}\mathbf{k}^2\right)\tilde{P} + \left(\hat{f}\mathbf{k}^2 - 2i\hat{q}\mathbf{k}P_S - 4i\hat{z}\mathbf{k}P_S^3 - M\omega^2\right)\tilde{U} = \tilde{E}. \quad (A.6b)$$

Hereinafter the following designations are introduced [31]:

$$\hat{c}\mathbf{k}^2 \stackrel{\text{def}}{=} c_{55}k_1^2 + 2c_{53}k_1 k_3 + c_{44}k_2^2 + c_{33}k_3^2, \quad (A.6c)$$

$$\hat{v}\mathbf{k}^4 \stackrel{\text{def}}{=} v_{3ij3lm}k_i k_j k_l k_m = v_{5151}k_1^4 + 2v_{4233}k_1^2 k_3^2 + v_{4242}k_2^4 + v_{3333}k_3^4, \quad (A.6d)$$

$$\hat{f}\boldsymbol{k}^2 \stackrel{\text{def}}{=} f_{55}k_1^2 + f_{44}k_2^2 + (f_{35} + f_{35})k_1 k_3 + f_{33}k_3^2 \quad (A.6e)$$

$$\hat{g}\boldsymbol{k}^2 \stackrel{\text{def}}{=} g_{55}k_1^2 + g_{44}k_2^2 + 2g_{35}k_1 k_3 + g_{33}k_3^2 + \frac{k_3^2}{\varepsilon_0 \varepsilon_b k^2}, \quad (A.6f)$$

$$\hat{q}\boldsymbol{k} \stackrel{\text{def}}{=} q_{53}k_1 + q_{33}k_3, \quad \hat{z}\boldsymbol{k} \stackrel{\text{def}}{=} z_{533}k_1 + z_{333}k_3, \quad (A.6g)$$

$$k^2 \stackrel{\text{def}}{=} k_1^2 + k_2^2 + k_3^2, \quad (A.6h)$$

$$\alpha_0 = \alpha^* + 3\beta^*(P_3^0)^2 + 5\gamma(P_3^0)^4 + 7\delta(P_3^0)^6. \quad (A.6i)$$

Note that $\alpha_0$ is always positive. The coefficients $\alpha^* = \alpha - 2(q_{53}u_{31}^0 + q_{33}u_{33}^0)$ and $\beta^* = \beta - 4(z_{533}u_{31}^0 + z_{333}u_{33}^0)$ renormalized by elastic strains are introduced.

From the singularity of the generalized susceptibility, corresponding to the condition $\Delta(\boldsymbol{k}, \omega) = 0$, follows the characteristic equation for the frequency $\omega(\boldsymbol{k})$:

$$(\mu\rho - M^2)\omega^4 + i(\Gamma\rho + \Lambda\mu)\omega^3 - C(\boldsymbol{k})\omega^2 - i\omega\left[\Gamma(\hat{v}\boldsymbol{k}^4 + \hat{c}\boldsymbol{k}^2) + \Lambda(\alpha_S + \hat{g}\boldsymbol{k}^2)\right] + B(\boldsymbol{k}) = 0. \quad (A.7a)$$

Hereinafter the functions $C(\boldsymbol{k})$ and $B(\boldsymbol{k})$ are used:

$$C(\boldsymbol{k}) = \alpha_0\rho + \Gamma\Lambda + (\hat{c}\boldsymbol{k}^2\mu - 2\hat{f}\boldsymbol{k}^2 M + \hat{g}\boldsymbol{k}^2\rho) + \mu\hat{v}\boldsymbol{k}^4, \quad (A.7b)$$

$$B(\boldsymbol{k}) = \alpha_0 \hat{c}\boldsymbol{k}^2 - 4(P_3^0)^2[\hat{q}\boldsymbol{k} + 2\hat{z}\boldsymbol{k}(P_3^0)^2]^2 + \hat{c}\boldsymbol{k}^2 \hat{g}\boldsymbol{k}^2 + \alpha_0\hat{v}\boldsymbol{k}^4 - (\hat{f}\boldsymbol{k}^2)^2 + \hat{g}\boldsymbol{k}^2\hat{v}\boldsymbol{k}^4. \quad (A.7c)$$

For negligibly small or absent damping (i.e., at $\Gamma = \Lambda = 0$), the solution of the biquadratic Eq.(A.7a) acquires a simple form:

$$\omega_{O,A}^2(\boldsymbol{k}) = \frac{C(\boldsymbol{k}) \pm \sqrt{C^2(\boldsymbol{k}) - 4(\mu\rho - M^2)B(\boldsymbol{k})}}{2(\mu\rho - M^2)}. \quad (A.8)$$

Dispersion relation (A.8) contains one optical (**O**) and one (**A**) phonon modes, which corresponds to the signs "+" and "−" before the radical, respectively.

In a particular case $k_3 = 0$, Eqs.(A.6) acquire the form:

$$\hat{c}\boldsymbol{k}^2 = c_{55}k_1^2 + c_{44}k_2^2 = c(\varphi)k_\perp^2, \quad c(\varphi) = c_{55}\cos^2\varphi + c_{44}\sin^2\varphi, \quad (A.9a)$$

$$\hat{v}\boldsymbol{k}^4 = v_{5151}k_1^4 + v_{4242}k_2^4 = v(\varphi)k_\perp^4, \quad v(\varphi) = v_{5151}\cos^4\varphi + v_{4242}\sin^4\varphi, \quad (A.9b)$$

$$\hat{f}\boldsymbol{k}^2 \stackrel{\text{def}}{=} f_{55}k_1^2 + f_{44}k_2^2 = f(\varphi)k_\perp^2, \quad f(\varphi) = f_{55}\cos^2\varphi + f_{44}\sin^2\varphi, \quad (A.9c)$$

$$\hat{g}\boldsymbol{k}^2 \stackrel{\text{def}}{=} g_{55}k_1^2 + g_{44}k_2^2 = g(\varphi)k_\perp^2, \quad g(\varphi) = g_{55}\cos^2\varphi + g_{44}\sin^2\varphi, \quad (A.9d)$$

$$\hat{q}\boldsymbol{k} = q_{53}\cos\varphi\, k_\perp, \quad \hat{z}\boldsymbol{k} = z_{533}\cos\varphi\, k_\perp, \quad (A.9e)$$

$$k^2 = k_1^2 + k_2^2 \equiv k_\perp^2, \quad \tan\varphi = \frac{k_1}{k_2}. \quad (A.9f)$$

Expression (A.7c) can be rewritten as:

$$B(\boldsymbol{k}) = (\alpha_0 + \hat{g}\boldsymbol{k}^2)(\hat{c}\boldsymbol{k}^2 + \hat{v}\boldsymbol{k}^4) - 4P_S^2(\hat{q}\boldsymbol{k} + 2\hat{z}\boldsymbol{k}P_S^2)^2 - (\hat{f}\boldsymbol{k}^2)^2 =$$

$$= [\alpha_0 + g(\varphi)k_\perp^2][c(\varphi)k_\perp^2 + v(\varphi)k_\perp^4] - 4(P_3^0)^2[q_{53} + 2(P_3^0)^2 z_{533}]^2\cos^2\varphi\, k_\perp^2 - [f(\varphi)\, k_\perp^2]^2 \quad (A.10)$$

The condition $B(\boldsymbol{k}) = 0$ could be written as

$$g(\varphi)v(\varphi)k_\perp^6 + \left(v(\varphi)\alpha_0 - f(\varphi)^2 + c(\varphi)g(\varphi)\right)k_\perp^4 + c(\varphi)\alpha_0 k_\perp^2 - 4(P_3^0)^2(q_{53} +$$

$$2(P_3^0)^2 z_{533})^2 \cos^2\varphi\, k_\perp^2 = 0. \quad (A.12)$$

From Eq.(A.12), the analytical expression for the field-dependent critical value $k_{cr}$, which corresponds to $B(\mathbf{k}) = 0$, is:

$$k_{cr}^2(\varphi) = \frac{f^2(\varphi) - f_{th}^2(\varphi) \pm \sqrt{[f^2(\varphi) - f_{th}^2(\varphi)]^2 - \xi_{sm}(\varphi)}}{2g(\varphi)v(\varphi)} \equiv$$

$$\frac{f^2(\varphi) - f_{th}^2(\varphi) \pm \sqrt{[f^2(\varphi) - f_{th}^2(\varphi) - \sqrt{\xi_{sm}(\varphi)}][f^2(\varphi) - f_{th}^2(\varphi) + \sqrt{\xi_{sm}(\varphi)}]}}{2g(\varphi)v(\varphi)}, \quad (A.13a)$$

where the value $f_{th}(\varphi)$ is an upper thermodynamic limit [10, 11, 19] of the flexoelectric coefficient

$$f_{th}^2(\varphi) = v(\varphi)\alpha_0 + c(\varphi)g(\varphi). \quad (A.13b)$$

The value $\xi_{sm}(\varphi)$ is introduced as

$$\xi_{sm}(\varphi) = 4g(\varphi)v(\varphi)[c(\varphi)\alpha_0 - 4(P_3^0)^2(q_{53} + 2(P_3^0)^2 z_{533})^2 \cos^2\varphi]. \quad (A.13c)$$

Assuming that $\xi_{sm} \geq 0$, which is the most common case for small external fields, because $g(\varphi) > 0$, $v(\varphi) > 0$, $c(\varphi) > 0$ and $\alpha_0 > 0$, the minimal and maximal critical values of $f(\varphi)$ is given by expression

$$f_{cr1,2}^2(\varphi) = f_{th}^2(\varphi) \mp \sqrt{\xi_{sm}(\varphi)}, \quad (A.13c)$$

As follows from Eq.(A.13a), the real wavevectors can exist only for $f^2(\varphi) < f_{cr1}^2(\varphi)$ and $f^2(\varphi) > f_{cr2}^2(\varphi)$. Since $f_{cr1}^2(\varphi) \leq f_{th}^2(\varphi) \leq f_{cr2}^2(\varphi)$, only the second case $f^2(\varphi) > f_{cr2}^2(\varphi)$ corresponds to $k_{cr}^2(\varphi) \geq 0$ and thus has physical sense.

Assuming that $\xi_{sm} < 0$, which is possible for small external fields, the solution $k_{cr}^2(\varphi) = \frac{f^2(\varphi) - f_{th}^2(\varphi) + \sqrt{[f^2(\varphi) - f_{th}^2(\varphi)]^2 - \xi_{sm}(\varphi)}}{2g(\varphi)v(\varphi)}$ exists for all values of $f^2$, and so the critical value $f_{cr}^2(\varphi) = 0$. Thus, the true critical value of the flexoelectric coupling coefficient is

$$f_{cr}^2(\varphi) = f_{th}^2(\varphi) + \sqrt{\xi_{sm}(\varphi)}, \quad \xi_{sm}(\varphi) \geq 0, \quad (A.14a)$$

$$f_{cr}^2(\varphi) = 0, \quad \xi_{sm}(\varphi) < 0. \quad (A.14b)$$

**Table AI.** LGD parameters for a bulk ferrielectric CuInP$_2$S$_6$ Helmholtz free energy $F(P_3, u_{ij})$ with $P_3$ and $u_{ij}$ as independent variables (collected from Refs. [37-42])

| Parameter (dimensionality) | Value |
|---|---|
| $\varepsilon_b$ | 9 |
| $\alpha_T$ (C$^{-2}$·m J/K) | $1.64067 \times 10^7$ |
| $T_{C,q}$ (K) | $T_C \cong 292.67$, $T_q \cong 50$ |
| $\beta$ (C$^{-4}$·m$^5$J) | $8.43 \times 10^{12}(1. - 0.00239\,T + 2.28 \times 10^{-6}T^2)$ |
| $\gamma$ (C$^{-6}$·m$^9$J) | $-1.67283 \times 10^{16}(1 - 0.00249\,T + 3.389 \times 10^{-6}\,T^2)$ |
| $\delta$ (C$^{-8}$·m$^{13}$J) | $9.824 \times 10^{18}(1. - 0.00127\,T + 4.0999 \times 10^{-6}T^2)$ |
| $q_{i3}$ (J C$^{-2}$·m) | $q_{13} = 1.4879 \times 10^{11}(1 - 0.00206\,T)$ $q_{23} = 1.0603 \times 10^{11}(1 - 0.00203\,T)$ $q_{33} = -4.0334 \times 10^{11}(1 - 0.00188\,T)$ $q_{53} = -7.3 * 10^{10}$ |
| $z_{i33}$ (C$^{-4}$·m$^5$J)$^*$ | $z_{133} = -1.414 \times 10^{14}(1 - 0.00099\,T)$ |

| | | |
|---|---|---|
| | | $z_{233} = -0.774 \times 10^{14}(1 - 0.00146\,T)$<br>$z_{333} = 1.181 \times 10^{14}(1 - 0.00699\,T)$<br>$z_{533} = 10^{13}$ |
| $c_{ij}$ (Pa) | | $c_{11} = 9.986 \times 10^{10}$, $c_{12} = 2.901 \times 10^{10}$, $c_{13} = -0.86 \times 10^9$, $c_{23} = -1.93 \times 10^9$, $c_{22} = 10.17 \times 10^{10}$, $c_{33} = 2.802 \times 10^{10}$, $c_{44} = 6.99 \times 10^9$, $c_{55} = 6.71 \times 10^9$, $c_{66} = 3.756 \times 10^{10}$ |
| $g_{3i3j}$ (J m³/C²)*** | | Estimated parameter, which has an order of $10^{-10}$, e.g., $g \cong (0.3 - 2.0) \times 10^{-9}$ |
| $f_{55}$ (V) | | 0-10 |
| $v_{311311}$ | | $3\,10^{-9}$ |
| $\Gamma$ (s m J/C²) | | ~$10^{-3}$ |
| $\mu$ (s² m J/C²) | | $8 \ast 10^{-14}$ |
| $\rho$ (kg/m³) | | 3427 |
| $M$ (s² J/(m C²)) | | $10^{-11}$ |

**Table AII**. LGD parameters for a bulk ferrielectric CuInP$_2$S$_6$ Hibbs free energy $G(P_3, \sigma_{ij})$ with $P_3$ and $\sigma_{ij}$ as independent variables (collected from Refs. [37-42])

| Designation | Units | Numerical value |
|---|---|---|
| $\varepsilon_b$ | Dimensionless | 9 |
| $\alpha_T$ | C⁻²·m J/K | $1.64067 \times 10^7$ |
| $T_C$ | K | 292.67 |
| $\beta$ | C⁻⁴·m⁵J | $3.148 \times 10^{12}$ |
| $\gamma$ | C⁻⁶·m⁹J | $-1.0776 \times 10^{16}$ |
| $\delta$ | C⁻⁸·m¹³J | $7.6318 \times 10^{18}$ |
| $Q_{i3}$ | C⁻²·m⁴ | $Q_{13} = 1.70136 - 0.00363\,T$, $Q_{23} = 1.13424 - 0.00242\,T$, $Q_{33} = -5.622 + 0.0105\,T$ |
| $Z_{i33}$ | C⁻⁴·m⁸ | $Z_{133} = -2059.65 + 0.8\,T$, $Z_{233} = -1211.26 + 0.45\,T$, $Z_{333} = 1381.37 - 12\,T$ |
| $s_{ij}$ | Pa⁻¹ | $s_{11} = 1.092 \times 10^{-11}$, $s_{12} = -0.311 \times 10^{-11}$, $s_{13} = +0.0120 \times 10^{-11}$, $s_{22} = 1.074 \times 10^{-11}$, $s_{23} = +0.0644 \times 10^{-11}$, $s_{33} = 3.574 \times 10^{-11}$, $s_{44} = 14.31 \times 10^{-11}$, $s_{55} = 14.90 \times 10^{-11}$, $s_{66} = 2.662 \times 10^{-11}$ |

### A3. Analytical Solution for the Eigen Frequency of the Flexoferrons

Following the seminal approach of Tang et al. [23], we consider the case when the mechanical force $N$ and the electric field $E$ in the right side of Eq.(A.3) are Langevin noise fields, which obey the fluctuation-dissipation theorem. Namely, their correlator (averaged over the statistical ensemble in the Fourier space) in the "quantum" case are

$$\langle \tilde{E}(\boldsymbol{k}, \omega)\tilde{E}^*(\boldsymbol{k}', \omega')\rangle = \frac{\Gamma}{(2\pi)^2} \hbar\omega \coth\left(\frac{\hbar\omega}{2k_B T}\right) \delta(\boldsymbol{k} - \boldsymbol{k}')\delta(\omega - \omega'), \qquad (A.15a)$$

$$\langle \tilde{N}(\boldsymbol{k}, \omega)\tilde{N}^*(\boldsymbol{k}', \omega')\rangle = \frac{\Lambda}{(2\pi)^2} \hbar\omega \coth\left(\frac{\hbar\omega}{2k_B T}\right) \delta(\boldsymbol{k} - \boldsymbol{k}')\delta(\omega - \omega'). \qquad (A.15b)$$

Also $\langle \tilde{E}(\mathbf{k}, \omega) \rangle = 0$ and $\langle \tilde{N}(\mathbf{k}, \omega) \rangle = 0$.

Hereinafter we neglect the damping of the elastic phonons, i.e., put $\Lambda = 0$, for the sake of simplicity. For $\Lambda = 0$ the linear polarization response can be found in the first order of the perturbation theory as shown in Ref.[31]. Following Ref.[31], the second-order correction in the case of quantum noise is

$$\langle p(\mathbf{x}, t) \rangle \cong -\Gamma \frac{\eta}{\alpha_0} \int_{-\infty}^{\infty} \frac{d\omega}{2\pi} \hbar\omega \coth\left(\frac{\hbar\omega}{2k_B T}\right) \int_{-\infty}^{\infty} \frac{d^3 k}{(2\pi)^3} |\tilde{\chi}(\mathbf{k}, \omega)|^2. \quad \text{(A.16a)}$$

where we used that $\tilde{\chi}(0,0) = \frac{1}{\alpha_0}$ and introduce the function $\eta = 3\beta^* P_3^0 + 10\gamma (P_3^0)^3 + 21\delta (P_3^0)^5$. Since $\hbar\omega \coth\left(\frac{\hbar\omega}{2k_B T}\right) \to 2k_B T$ in the classical white noise limit, $\frac{\hbar\omega}{k_B T} \to 0$, Eq.(A.16a) transforms to the classical limit:

$$\langle p(\mathbf{x}, t) \rangle \cong -2k_B T \Gamma \frac{\eta}{\alpha_0} \int_{-\infty}^{\infty} \frac{d^3 k}{(2\pi)^3} \int_{-\infty}^{\infty} \frac{d\omega}{2\pi} |\tilde{\chi}(\mathbf{k}, \omega)|^2, \quad \text{(A.16b)}$$

The approximate expression for $\langle p \rangle$ is:

$$\langle p \rangle \approx \int_{-\infty}^{\infty} \frac{d^3 k}{(2\pi)^3} \left[ \coth\left(\frac{\hbar\omega_A(\mathbf{k})}{2k_B T}\right) \delta p_A(\mathbf{k}) + \coth\left(\frac{\hbar\omega_O(\mathbf{k})}{2k_B T}\right) \delta p_O(\mathbf{k}) \right], \quad \text{(A.17a)}$$

where the "spectral densities" of flexoferrons are

$$\delta p_O(\mathbf{k}) \approx \frac{-\hbar}{2(\mu\rho - M^2)} \frac{\eta}{\alpha_0} \left| \frac{\hat{v} k^4 + \hat{c} k^2 - \rho \omega_O^2(\mathbf{k})}{\left(\omega_A^2(\mathbf{k}) - \omega_O^2(\mathbf{k})\right) \omega_O(\mathbf{k})} \right|, \quad \text{(A.17b)}$$

$$\delta p_A(\mathbf{k}) \approx \begin{cases} \frac{-\hbar}{2(\mu\rho - M^2)} \frac{\eta}{\alpha_0} \left| \frac{\hat{v} k^4 + \hat{c} k^2 - \rho \omega_A^2(\mathbf{k})}{\left(\omega_O^2(\mathbf{k}) - \omega_A^2(\mathbf{k})\right) \omega_A(\mathbf{k})} \right|, & \omega_A^2(\mathbf{k}) > 0, \\ \frac{-\hbar}{2(\mu\rho - M^2)} \frac{\eta}{\alpha_0} \left| \frac{\hat{v} k^4 + \hat{c} k^2 - \rho \omega_A^2(\mathbf{k})}{\left(\omega_O^2(\mathbf{k}) - \omega_A^2(\mathbf{k})\right) \omega_A(\mathbf{k})} \right| \frac{\Gamma}{\sqrt{\Gamma^2 - 4\mu^2 \omega_A^2(\mathbf{k})}}, & \omega_A^2(\mathbf{k}) < 0. \end{cases} \quad \text{(A.17c)}$$

Note that $\omega_O^2(\mathbf{k})$ is always positive, meanwhile the case $\omega_A^2(\mathbf{k}) < 0$ is possible for high flexoelectric coefficients and/or high electric fields close to coercive value. To derive Eqs.(A.17) we used the following integrals:

$$\int_{-\infty}^{\infty} \frac{(\omega^2 - \omega_0^2) d\omega}{[\mu(\omega_p^2 - \omega^2)]^2 + (\omega \Gamma)^2} = \begin{cases} \frac{\pi(\omega_p^2 - \omega_0^2)}{\mu \Gamma \omega_p^2}, & \omega_p^2 > 0, \\ \frac{\pi(\omega_p^2 + \omega_0^2)}{\mu \omega_p^2 \sqrt{\Gamma^2 - 4\mu^2 \omega_p^2}}, & \omega_p^2 < 0. \end{cases} \quad \text{(A.18a)}$$

$$\int_{-\infty}^{\infty} \frac{\frac{1}{2\pi} \hbar\omega \coth\left(\frac{\hbar\omega}{2k_B T}\right) d\omega}{\left[\frac{(\mu\rho - M^2)(\omega_O^2 - \omega^2)(\omega_A^2 - \omega^2)}{\rho\omega^2 - ck^2 - vk^4}\right]^2 + (\omega \Gamma)^2} \approx \int_{-\infty}^{\infty} \frac{\frac{1}{2\pi}\left\{\omega^2 - \frac{ck^2 + vk^4}{\rho}\right\}\left\{\omega_A^2 - \frac{ck^2 + vk^4}{\rho}\right\} \hbar\omega_A \coth\left(\frac{\hbar\omega_A}{2k_B T}\right) d\omega}{\left[\left(\mu - \frac{M^2}{\rho}\right)(\omega_O^2 - \omega_A^2)(\omega_A^2 - \omega^2)\right]^2 + \left(\left\{\omega_A^2 - \frac{ck^2 + vk^4}{\rho}\right\}\omega\Gamma\right)^2} +$$

$$\int_{-\infty}^{\infty} \frac{\frac{1}{2\pi}\left\{\omega^2 - \frac{ck^2 + vk^4}{\rho}\right\}\left\{\omega_O^2 - \frac{ck^2 + vk^4}{\rho}\right\} \hbar\omega_O \coth\left(\frac{\hbar\omega_O}{2k_B T}\right) d\omega}{\left[\left(\mu - \frac{M^2}{\rho}\right)(\omega^2 - \omega_O^2)(\omega_O^2 - \omega_A^2)\right]^2 + \left(\left\{\omega_O^2 - \frac{ck^2 + vk^4}{\rho}\right\}\omega\Gamma\right)^2}, \quad \text{(A.18b)}$$

# Appendix B. Additional Figures

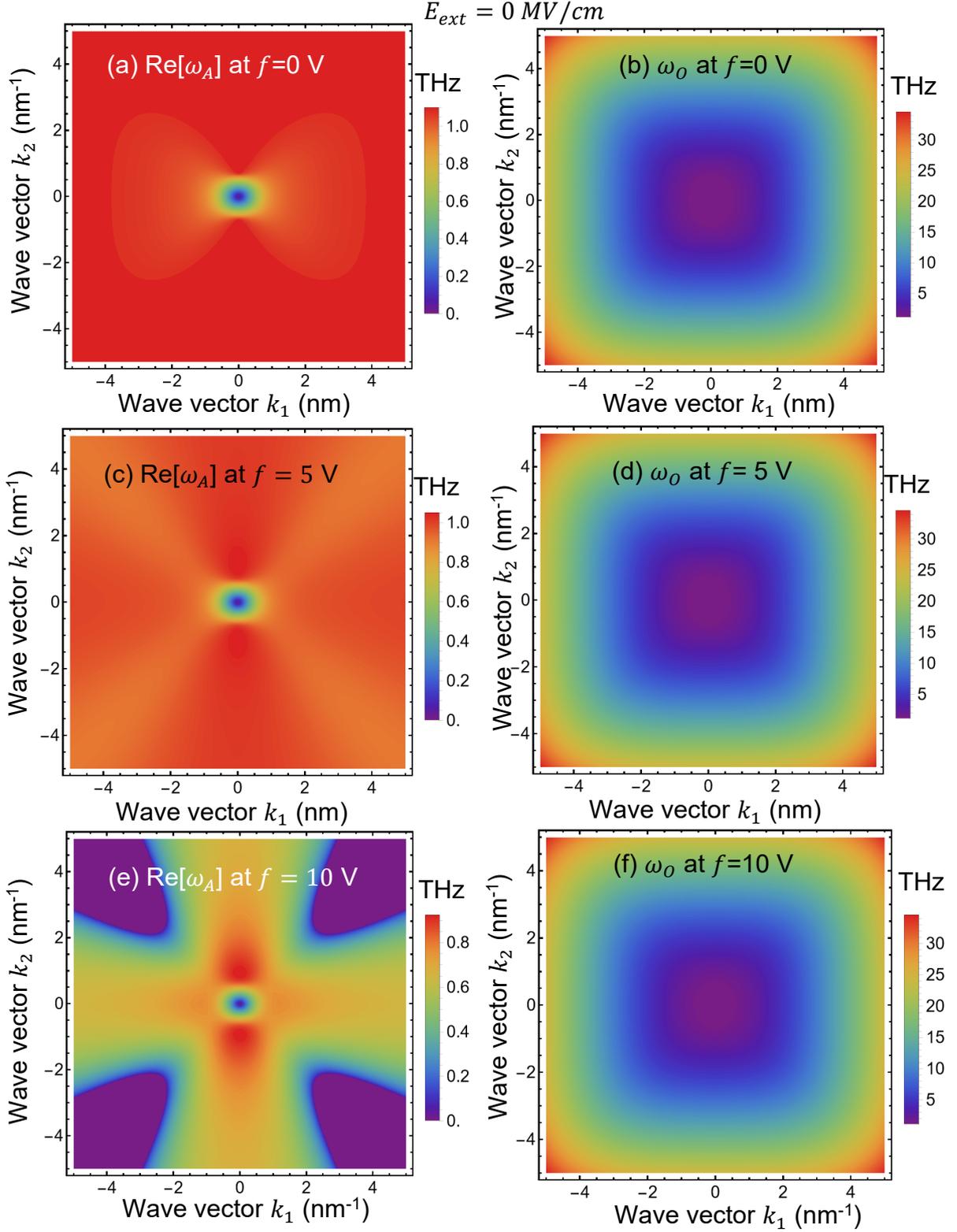

**FIGURE B1.** The real part of the acoustic phonon frequency, $Re[\omega_A]$, **(a, c, e)**; and the optical phonon frequency, $\omega_O$ **(b, d, f)**, as the function of the wavevector components $k_1$ and $k_2$ calculated for different values of the flexoelectric coefficient $f$=0 **(a, b)**, 5 **(c, d)** and 10 V **(e, f)**. The electric field is zero. The damping is absent ($\Gamma = 0, \Lambda = 0$), $k_3 = 0$ and $T = 10$ K for all plots. Material parameters of $CuInP_2S_6$ are listed in **Table AI**.

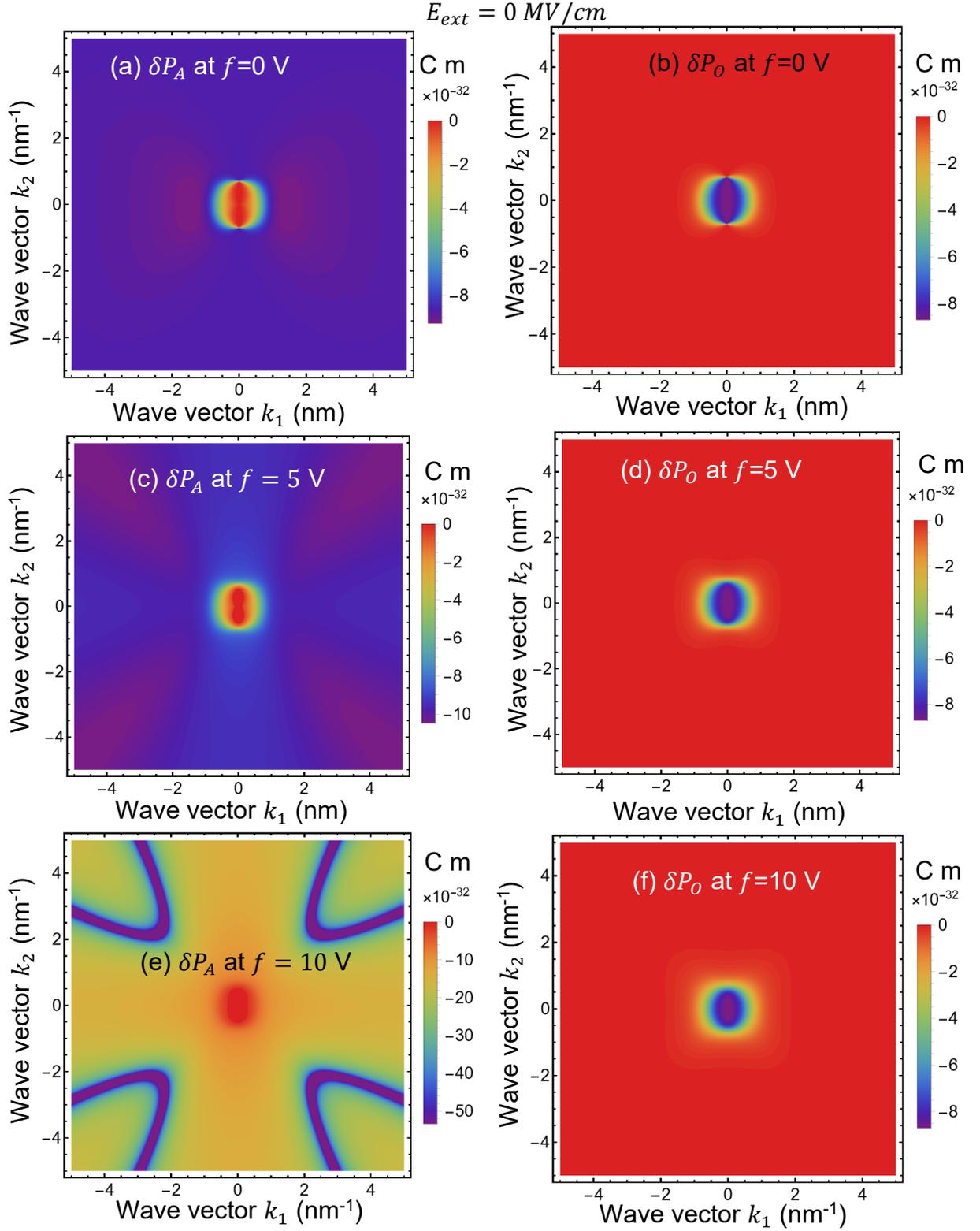

**FIGURE B2.** The spectral densities of the acoustic and optic flexoferrons, $\delta p_A$ (a, c, e) and $\delta p_O$ (b, d, f), as the function of the wavevector components $k_1$ and $k_2$ for $k_3 = 0$, calculated for different values of the flexoelectric coefficient $f$=0 (a, b), 5 (c, d) and 10 V (e, f). The electric field is zero. The damping is absent ($\Gamma = 0, \Lambda = 0$), $k_3 = 0$ and $T = 10$ K for all plots. Material parameters of CuInP$_2$S$_6$ are listed in **Table AI**.

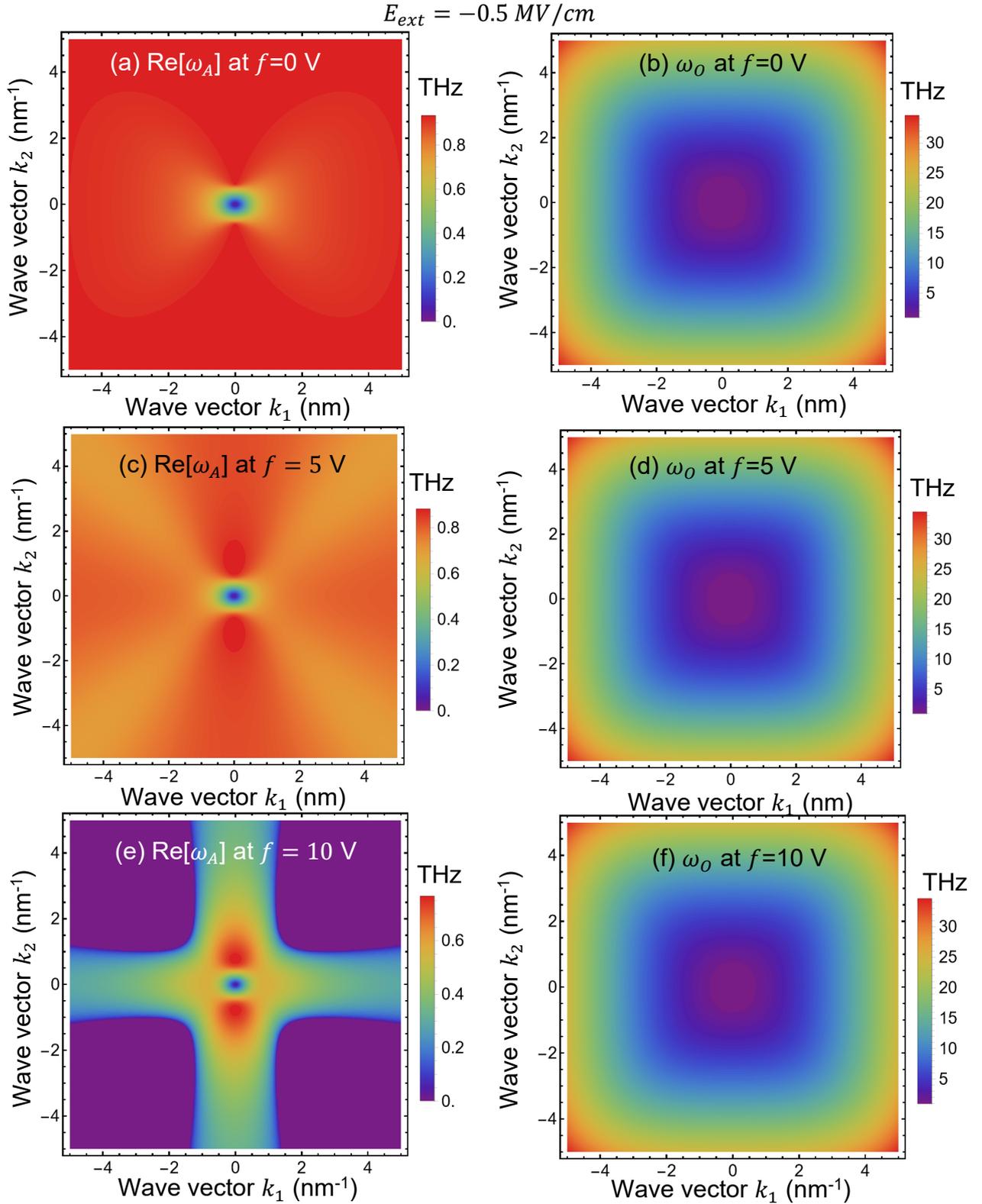

**FIGURE B3.** The real part of the acoustic phonon mode frequency, $Re[\omega_A]$, **(a, c, e)**; and the optical phonon frequency, $\omega_O$ **(b, d, f)**, as the function of the wavevector components $k_1$ and $k_2$ calculated for different values of the flexoelectric coefficient $f$=0 **(a, b)**, 5 **(c, d)** and 10 V **(e, f)**. The electric field is - 0.5 MV/cm. The damping is absent ($\Gamma = 0, \Lambda = 0$), $k_3 = 0$ and $T = 10$ K for all plots. Material parameters of CuInP$_2$S$_6$ are listed in **Table AI**.

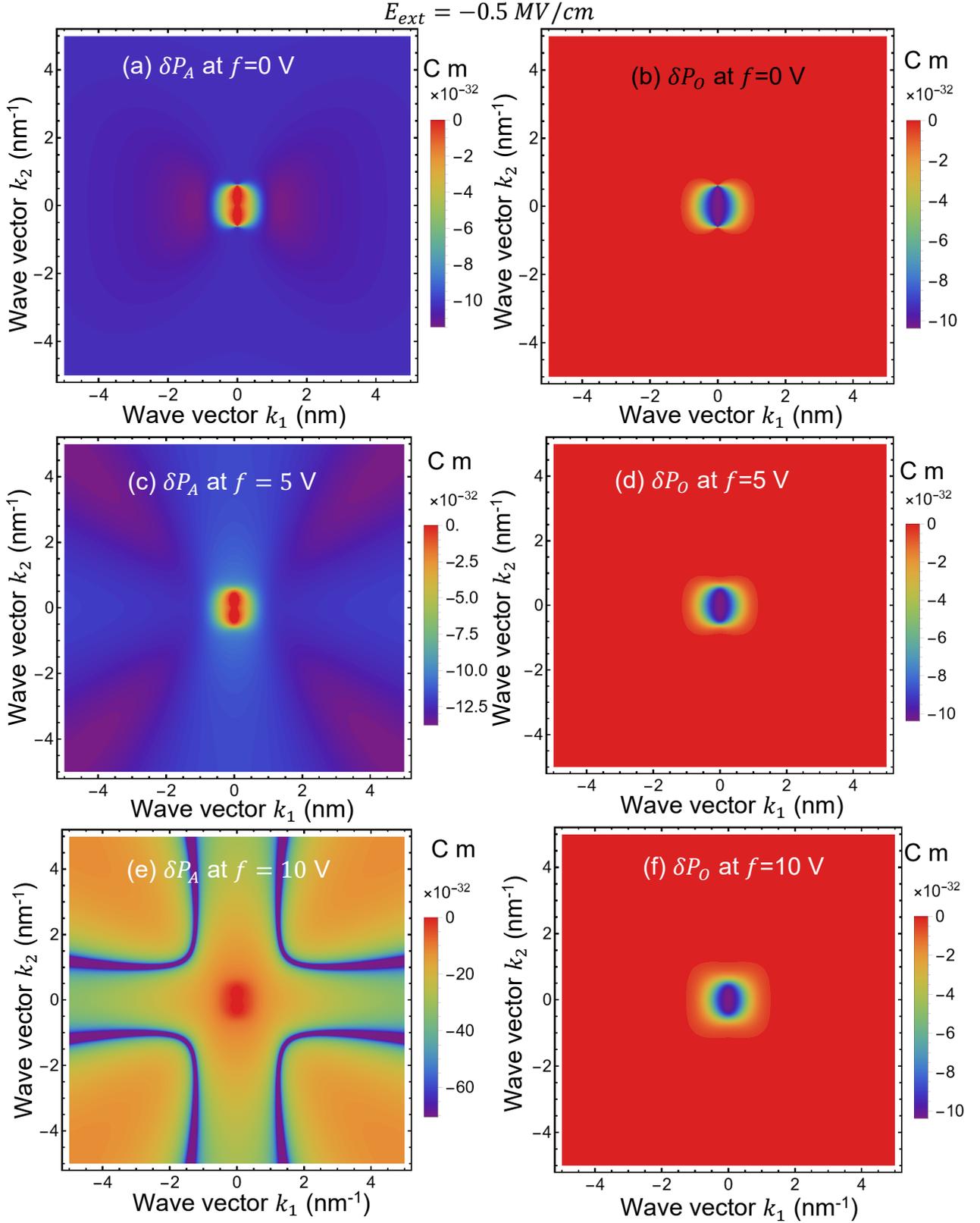

**FIGURE B4.** The spectral densities of the acoustic and optical flexoferrons, $\delta p_A$ **(a, c, e)** and $\delta p_O$ **(b, d, f)**, as the function of the wavevector components $k_1$ and $k_2$ calculated for different values of the flexoelectric coefficient $f$=0 **(a, b)**, 5 **(c, d)** and 10 V **(e, f)**. The electric field is - 0.5 MV/cm. The damping is absent ($\Gamma = 0, \Lambda = 0$), $k_3 = 0$ and $T = 10$ K for all plots. Material parameters of CuInP$_2$S$_6$ are listed in **Table AI**.